\documentclass{article}
\usepackage{amsmath}
\usepackage{graphicx}
\usepackage{amsfonts}
\usepackage{amssymb}
\renewcommand{\theequation}{\arabic{section}.\arabic{equation}}

\begin{document}

\title{Asymptotic behavior of large polygonal Wilson loops in confining gauge theories}
\author{P.V. Pobylitsa \\
\emph{Petersburg Nuclear Physics Institute}\\
\emph{Gatchina, 188300, St. Petersburg, Russia}}
\date{}
\maketitle

\textbf{Abstract}

In the framework of effective string theory (EST), the asymptotic behavior of
a large Wilson loop in confining gauge theories can be expressed via Laplace
determinant with Dirichlet boundary condition on the Wilson contour. For
a general polygonal region, Laplace determinant can be computed using the conformal
anomaly and Schwarz-Christoffel transformation. One can construct ratios of
polygonal Wilson loops whose large-size limit can be expressed via computable
Laplace determinants and is independent of the (confining) gauge group. These
ratios are computed for hexagon polygons both in EST and by Monte Carlo (MC)
lattice simulations for the tree-dimensional lattice $\mathbb{Z}_{2}$ gauge
theory (dual to Ising model) near its critical point. For large hexagon Wilson
loops a perfect agreement is observed between the asymptotic EST expressions
and the lattice MC results.

\renewcommand{\arraystretch}{1.2} 

\section{Introduction}

\setcounter{equation}{0} 

\subsection{Large Wilson loops and effective string theory}

Historically, Wilson loops
\begin{equation}
W\left(  C\right)  =\left\langle \mathrm{Tr}\,\exp\left[  i\oint_{C}dx^{\mu
}A_{\mu}\left(  x\right)  \right]  \right\rangle
\end{equation}
attracted attention in the context of the problem of the heavy-quark
confinement although the full spectrum of problems where Wilson loops appear
is much wider. Gauge theories whose Wilson loops (at least in some
representations of the gauge group) obey area law \cite{Wilson-74}
\begin{equation}
\lim_{\left|  C\right|  \rightarrow\infty}\frac{1}{S\left(  C\right)  }\ln
W\left(  C\right)  =-\sigma\neq0 \label{Wilson-area-law}
\end{equation}
(with $S\left(  C\right)$ being the area of the surface spanned on contour $C$ and with
$\sigma$ interpreted as the string tension) are often briefly called confining
gauge theories. However, the real problem of confinement is more general and
can be reduced to area law (\ref{Wilson-area-law}) only under certain assumptions: heavy-quark limit,
ignoring exceptional cases when the confining potential is not asymptotically
linear and some other subtleties.

In eq.~(\ref{Wilson-area-law}), the non-formal notation
\begin{equation}
\left|  C\right|  \rightarrow\infty
\end{equation}
is used for the large-size limit of contour $C$, implying that the contour
grows more or less uniformly in all directions. For brevity, we will often use
expression \emph{large Wilson loop}, keeping in mind the large-size limit
$\left|  C\right|  \rightarrow\infty$ and not the value of the Wilson loop
which is exponentially suppressed according to eq. ~(\ref{Wilson-area-law}).
Typically, limit $\left|  C\right|  \rightarrow\infty$ assumes that area
$S\left(  C\right)  $ and perimeter $L\left(  C\right)  $ of contour $C$ scale
as
\begin{equation}
S\left(  C\right)  \sim L^{2}\left(  C\right)  \rightarrow\infty\,.
\label{S-C-orders}
\end{equation}
In case of area law (\ref{Wilson-area-law}) condition (\ref{S-C-orders}) can
be relaxed. On the other hand, the analysis of higher-order terms of the
large-size expansion of $W\left(  C\right)  $ may require a more careful
specification of details of the large-size limit (e.g. the uniform rescaling
of contour $C$ by a large factor).

The modern understanding of confining gauge theories is based on the idea that
area law is a manifestation of a more fundamental phenomenon: effective-string
formation (ESF) between static color charges separated by a large distance.
The concept of ESF has a long history full of imprecise qualitative ideas,
heuristic arguments and models without a solid theoretical
basis. But finally it became clear that ESF is a phenomenon which allows for a
systematic theoretical treatment. The phenomenon of ESF has many features
similar to spontaneous breakdown of continuous symmetries. In particular, the
large-distance (and low-energy) behavior of many quantities (including large
Wilson loops) can be described by an effective string action that contains an
infinite series of local terms ordered according to the counting of gradients
similarly to the method of effective actions describing Goldstone modes in
theories with a spontaneous breakdown of continuous symmetries. This approach
to the description of ESF based on the effective action and on the large-size
(or small-energy) expansion is usually called effective string theory (EST).

EST occupies a rather peculiar place in the large zoo of string-based theories:

1) EST is a systematic theory based on the expansion in a small parameter,
inverse size of the string (or small energy of string excitations).

2) EST is not a fundamental string theory. It is an effective theory which can
be applied only in the large-size limit.

The early time of EST was intertwined with the history of other
stringy models and theories for hadrons. As for the problem of confinement,
the string interpretation was emphasized by K. Wilson \cite{Wilson-74}
combining intuitive arguments with the properties of the strong coupling
expansion in lattice gauge theories. G.~'t~Hooft \cite{tHooft-1978},
\cite{tHooft-1979} constructed a classification of states of non-abelian gauge
theories in a finite periodic box. This classification combined with the limit
of large volume puts the concept of \emph{closed} periodic strings in
confining gauge theories on a solid theoretical ground.  \emph{Open} confining
strings are associated with pairs of external static color charges and are
described by Wilson loops. EST describes open and closed strings using the
same effective action (up to extra boundary terms which are essential for
higher-order corrections to Wilson loops and to the spectrum of open
strings). M.~L\"{u}scher, G.~M\"{u}nster, K.~Symanzik and P.~Weisz
\cite{Luscher:1980fr}-\cite{Luscher:1980iy} made several crucial steps on the
way from naive string models towards EST as a systematic effective theory. The
subsequent theoretical work in EST went in various directions including

-- derivation of general constraints on terms appearing in the action of EST
\cite{Luscher:2004ib}-\cite{BCGMP-12},

-- computation of loop corrections in EST for rectangular Wilson loops
\cite{BCGMP-12}, \cite{Filk-preprint}-\cite{Billo:2011fd} and for other
closely related quantities like correlation functions of Polyakov loops and
spectra of closed and open strings \cite{Luscher:2004ib},
\cite{Aharony:2009gg}-\cite{AK-2013},

-- analysis of string finite-width effects \cite{Luscher:1980iy},
\cite{Gliozzi:2010zt}, \cite{Meyer-2010}.

The predictions of EST have been successfully verified by many lattice MC
tests (see review \cite{BM-16}, recent publications \cite{BCGMP-12},
\cite{Athenodorou:2010cs}-\cite{AT-2016-b} and references therein).

One should keep in mind that the stringy interpretation for Wilson loops appears not
only in the context of the large-size limit but also in other approaches:
$1/N$-expansion \cite{Hooft-1974}-\cite{Migdal-1984}, large-$D$ limit ($D$
being the space-time dimension) \cite{Alvarez}-\cite{Makeenko-2012-a}, Regge
limit \cite{Makeenko-2012-a}-\cite{Makeenko-2012-b}. Wilson loops attract
much attention in superconformal theories (like 4D $\mathcal{N}=4$
supersymmetric Yang-Mills theory), especially in the combination with the
large-$N$ limit and with the AdS/CFT\ correspondence \cite{Maldacena-1998}
-\cite{ESZ-2000}, including the case of null polygonal Wilson loops
\cite{AM-2007}-\cite{BDM-2014} and tests of EST in holographic backgrounds
\cite{Aharony:2009gg}, \cite{Aharony:2010cx}. But these interesting fields of
research are beyond the scope of this work where EST is understood as an
effective theory relevant for the construction of the large-size expansion in
confining gauge theories with area law (\ref{Wilson-area-law}) without
assuming extra small parameters or extra symmetries.

\subsection{Reduction of large Wilson loops to Laplace determinants in EST}

EST is an effective theory. Its power is limited. For example, string tension
$\sigma$ appearing in area law (\ref{Wilson-area-law}) is determined by
microscopic gauge theory (MGT) and cannot be computed in EST. Therefore EST
provides incomplete information about large-size expansion of \emph{single
Wilson loops}. However, the large-size limit of certain \emph{combinations of
several Wilson loops} is completely computable in EST.

Let us take a set of \emph{smooth} flat contours $C_{i}$ ($1\leq i\leq n$).
Let us assign some number $m_{i}$ to each contour $C_{i}$ so that perimeters
$L\left(  C_{i}\right)  $ and areas $S\left(  C_{i}\right)  $ of these
contours obey constraints
\begin{equation}
\sum_{i=1}^{n}m_{i}=0\,, \label{balance-vertices-0}
\end{equation}
\begin{equation}
\sum_{i=1}^{n}m_{i}L\left(  C_{i}\right)  =0\,, \label{balance-perimeter-0}
\end{equation}
\begin{equation}
\sum_{i=1}^{n}m_{i}S\left(  C_{i}\right)  =0\,. \label{balance-area-0}
\end{equation}
Now let us rescale each contour $C_{i}\rightarrow\lambda C_{i}$ by a large
common factor $\lambda$. Then EST\ leads to the following expression for the
large-size limit:

\begin{equation}
\lim_{\lambda\rightarrow\infty}\prod_{i=1}^{n}\left[  W\left(  \lambda
C_{i}\right)  \right]  ^{m_{i}}=\left\{  \prod_{i=1}^{n}\left\{
\mathrm{Det}_{\zeta}\left[  -\Delta\left(  C_{i}\right)  \right]  \right\}
^{m_{i}}\right\}  ^{-(D-2)/2}\,. \label{W-product-det-product-0}
\end{equation}
Here $\Delta\left(  C_{i}\right)  $ is Laplace operator defined in the
two-dimensional region bounded by contour $C_{i}$ with Dirichlet boundary
condition. $D$ is the dimension of space-time. Notation $\mathrm{Det}_{\zeta
}\left[  -\Delta\left(  C_{i}\right)  \right]  $ stands for the determinant of
operator $-\Delta\left(  C_{i}\right)  $ in $\zeta$-regularization scheme (see
sec.~\ref{section-Laplace-determinants}). In this paper we usually assume
$\zeta$-regularization for Laplace determinants although in many (but not in
all) formulas one has a wider freedom of choice for the renormalization scheme.

Constraints (\ref{balance-vertices-0}) -- (\ref{balance-area-0}) are imposed
in order to cancel those contributions to the large-size expansion which are
controlled by MGT\ but not by EST. For example, condition
(\ref{balance-area-0}) cancels the $\sigma$ dependence in the product on the
LHS of (\ref{W-product-det-product-0}).

Asymptotic formula (\ref{W-product-det-product-0}) is a direct consequence of
the results obtained in ref.~\cite{Luscher:1980fr}. The derivation of eq.
(\ref{W-product-det-product-0}) is briefly discussed in sec.
\ref{section-Wilson-loops-EST}. The basic idea is that in the limit of large
size, Wilson loop $W\left(  C\right)  $ can be approximated by a functional
integral over surfaces bounded by contour $C$. The large size of the contour
justifies the steepest-decent expansion of this functional integral so that
the problem reduces to finding the minimum of the action (minimal surface
bounded by contour $C$) and to computing the determinant of the quadratic part
of the action for fluctuations near this minimal surface. For flat
contours $C$ the contribution of fluctuations is described by the determinant
of Laplace operator $\Delta\left(  C\right)  $ acting in the region bounded by
$C$ (with Dirichlet boundary condition). In $D$-dimensional space-time one has
$D-2$ transverse directions for the fluctuations of the string surface which
explains power $-(D-2)/2$ of Laplace determinant on the RHS\ of
(\ref{W-product-det-product-0}).

A remarkable feature of asymptotic formula (\ref{W-product-det-product-0}) is
that the RHS is independent of the gauge group (as long as we deal with a
confining gauge theory and ESF stands behind the confinement) and the dependence
on space-time dimension reduces to power $-\left(  D-2\right)  /2$.

\subsection{Case of polygonal Wilson loops}

The above discussion of asymptotic formula (\ref{W-product-det-product-0})
assumed that all contours $C_{i}$ are smooth. But when it comes to lattice
tests of EST, one has to deal with polygonal contours $C_{i}$.
Eq.~(\ref{W-product-det-product-0}) still works for polygonal contours $C_{i}$ if
one imposes an extra condition in addition to constraints
(\ref{balance-vertices-0}) -- (\ref{balance-area-0}). This extra constraint is
needed because some properties of Wilson loops near the vertices of polygons
are controlled by MGT and are not computable in EST: each vertex contributes an
EST-non-controllable factor depending on the vertex angle. In order to cancel
these factors on the LHS of (\ref{W-product-det-product-0}) let us choose some
set of angle values $\left\{  \theta_{a}\right\}  $ and let us distribute the
angles from this common set $\left\{  \theta_{a}\right\}  $ between different
polygons $C_{i}$ in a ``balanced way''. Let $p_{ia}$ be the number of
occurrences of angle value $\theta_{a}$ among vertices of polygon $C_{i}$ (so
that possible values are $p_{ia}=0,1,2,\ldots$). Then the vertex balance
constraint reads
\begin{equation}
\sum_{i=1}^{n}m_{i}p_{ia}=0\,. \label{balance-angles-0}
\end{equation}
If this condition is added to constraints (\ref{balance-vertices-0}) --
(\ref{balance-area-0}) then one can use asymptotic formula
(\ref{W-product-det-product-0}) for polygonal contours. In case of flat
polygonal contours condition (\ref{balance-vertices-0}) can be derived from
condition (\ref{balance-angles-0}) (see sec.
\ref{dependent-constraint-section}). Therefore for the validity of asymptotic
formula (\ref{W-product-det-product-0}) it is sufficient to impose only
constraints (\ref{balance-perimeter-0}), (\ref{balance-area-0}) and
(\ref{balance-angles-0}).

If one omits area constraint (\ref{balance-area-0}) but still keeps
constraints (\ref{balance-perimeter-0}), (\ref{balance-angles-0}) then eq.
(\ref{W-product-det-product-0}) can be generalized to the form (see sec.
\ref{derivation-asymptotic-formula-section})
\begin{align}
&  \left[  \exp\left(  \lambda^{2}\sigma\sum_{i=1}^{n}m_{i}S\left(
C_{i}\right)  \right)  \right]  \prod_{i=1}^{n}\left[  W\left(  \lambda
C_{i}\right)  \right]  ^{m_{i}}\nonumber\\
&  \overset{\lambda\rightarrow\infty}{=}\left\{  \prod_{i=1}^{n}\left\{
\mathrm{Det}_{\zeta}\left[  -\Delta\left(  C_{i}\right)  \right]  \right\}
^{m_{i}}\right\}  ^{-(D-2)/2}\left[  1+O\left(  \lambda^{-2}\right)  \right]
\,. \label{W-product-det-product-1}
\end{align}

Our guiding principle for building the product of Wilson loops on the LHS of
eq.~(\ref{W-product-det-product-0}) was to make such a combination of Wilson
loops that its limit $\lambda\rightarrow\infty$ is completely determined by
EST and is independent of the details of the MGT. Eq.
(\ref{W-product-det-product-1}) is written for a relaxed set of constraints:
we have omitted condition (\ref{balance-area-0}). The price is that relation
(\ref{W-product-det-product-1}) contains string tension $\sigma$ which cannot
be computed in EST and is determined by MGT. Note that parameter $\sigma$
appearing in (\ref{W-product-det-product-1}) is the same as in area law
(\ref{Wilson-area-law}). This statement is nontrivial because it assumes that
Laplace determinants on the RHS of eq.~(\ref{W-product-det-product-1}) are taken
in the $\zeta$-regularization. In fact, what is important is not $\zeta
$-regularization itself but its property that power divergences vanish in
$\zeta$-regularization (similarly to the dimensional regularization).

\subsection{Large-size expansion for single Wilson loops}

The above presentation (\ref{W-product-det-product-0}),
(\ref{W-product-det-product-1}) of EST results in terms of \emph{products of
Wilson loops}, whose large-size limit is either completely or almost
completely under control of EST, is determined by our final aim: lattice MC
tests of EST for polygonal Wilson loops. But for the derivation of asymptotic
relations (\ref{W-product-det-product-0}), (\ref{W-product-det-product-1}) one
needs the large-$\lambda$ expansion for a \emph{single} Wilson loop:
\begin{equation}
\ln W\left(  \lambda C\right)  \overset{\lambda\rightarrow\infty}{=}f_{\ln
}\left(  C\right)  \ln\lambda+\sum_{k\geq-2}\lambda^{-k}f_{k}\left(  C\right)
\,. \label{W-lambda-C-general-expansion-0}
\end{equation}
Here $f_{\ln}\left(  C\right)  $ and $f_{k}\left(  C\right)  $ are some
functionals of contour $C$. These functionals are partly computable in EST but
still are partly dependent on MGT. Large-$\lambda$ expansion
(\ref{W-lambda-C-general-expansion-0}) is discussed in detail in sec.
\ref{EST-one-loop-section}. At the moment it is important that the MGT
sensitivity of expansion (\ref{W-lambda-C-general-expansion-0}) cancels
completely in eq.~(\ref{W-product-det-product-0}) and almost completely (up to
the $\sigma$ dependence) in eq.~(\ref{W-product-det-product-1}) if one imposes
the corresponding set of constraints (\ref{balance-vertices-0}) --
(\ref{balance-area-0}), (\ref{balance-angles-0}).

The formal mathematical structure of large-$\lambda$ expansion
(\ref{W-lambda-C-general-expansion-0}) is rather general. We have an
asymptotic power series with an extra logarithmic term. Functionals
$f_{k}\left(  C\right)  $ have obvious properties
\begin{equation}
f_{k}\left(  \lambda C\right)  =\lambda^{-k}f_{k}\left(  C\right)
\quad\left(  k\neq0\right)  \,, \label{f-n-scaling}
\end{equation}
\begin{equation}
f_{0}\left(  \lambda C\right)  =f_{0}\left(  C\right)  +f_{\ln}\left(
C\right)  \ln\lambda\,,
\end{equation}
\begin{equation}
f_{\ln}\left(  \lambda C\right)  =f_{\ln}\left(  C\right)  \,.
\end{equation}
Expansion (\ref{W-lambda-C-general-expansion-0}) represents the current
understanding of EST and generalizes the results for loop corrections in EST
which have been computed or at least studied. In principle, apart from the
logarithmic term $f_{\ln}\left(  C\right)  \ln\lambda$, in higher of orders of
this expansion other ``non-analytical'' terms are not excluded.

The role of constraints (\ref{balance-vertices-0}) -- (\ref{balance-area-0}),
(\ref{balance-angles-0}) assumed for the validity of eq.
(\ref{W-product-det-product-0}) is to cancel the contribution of $\lambda
$-growing terms to eq.~(\ref{W-product-det-product-0})
\begin{equation}
\sum_{i=1}^{n}m_{i}f_{k}\left(  C_{i}\right)  \,=0\quad\left(  k=-1,-2,\ln
\right)  \,.
\end{equation}
Then eq.~(\ref{W-product-det-product-0}) becomes equivalent to
\begin{equation}
\sum_{i=1}^{n}m_{i}f_{0}\left(  C_{i}\right)  \,=-\frac{D-2}{2}\sum_{i=1}
^{n}m_{i}\ln\mathrm{Det}_{\zeta}\left[  -\Delta\left(  C_{i}\right)  \right]
\,.
\end{equation}
Explicit EST expressions for functionals $f_{k}\left(  C_{i}\right)$ can be
found in sec.~\ref{EST-one-loop-section}. Note that
\begin{equation}
f_{1}\left(  C\right)  =0 \label{f-1-zero}
\end{equation}
(see sec.~\ref{higher-orders-lambda-section}). This explains the $O\left(
\lambda^{-2}\right)  $ correction on the RHS of eq.
(\ref{W-product-det-product-1}) coming from the term $\lambda^{-2}f_{2}\left(
C\right)  $ of expansion (\ref{W-lambda-C-general-expansion-0}).

\subsection{Lattice tests of effective string theory}

Nowadays EST has the status of a mature theory. However, EST is an effective
theory, EST\ can be used only in cases when the phenomenon of ESF\ occurs. But
EST cannot tell us whether we really have ESF\ in a given microscopic gauge
theory or not. Also keep in mind that ESF is not the only possible mechanism
of confinement. Therefore lattice MC tests of EST play an important role.
Lattice MC simulations are also helpful for testing the internal kitchen of
EST because some arguments used in EST (e.g. some constraints on higher-order
terms of the effective action or the assumption that in the large-size limit
we effectively deal with a pure bosonic string without extra fields living on
the string world sheet) cannot be considered as rigorous theorems. Last but
not least: inconsistencies observed in lattice MC simulations may trigger
detecting errors in EST theoretical calculations as it happened with the
two-loop EST correction $f_{2}\left(  C\right)  \,$ for rectangular Wilson
loops (see sec.~\ref{higher-orders-lambda-section}).

Anyway, EST has successfully passed through many lattice MC tests. These tests
go far beyond Wilson loops and also include such quantities as correlation
functions of Polyakov loops, the confining potential, the spectrum of closed
and open strings, etc. The reader interested in general lattice tests of EST
beyond the special case of Wilson loops (which make the subject of this paper)
is referred to recent publications \cite{BM-16}, \cite{AT-2016-a}, \cite{AT-2016-b} and to
references therein.

In case of Wilson loops, MC tests of EST asymptotic formulas
(\ref{W-product-det-product-0}), (\ref{W-product-det-product-1}) have much
freedom of choice:

-- gauge group,

-- space-time dimension,

-- maximal order of EST corrections taken into account,

-- shape of the Wilson contour.

For \emph{rectangular} Wilson loops, EST has been successfully checked by
lattice MC tests with high accuracy resolving higher-order corrections of the
large-$\lambda$ expansion. In particular, in case of 3D $\mathbb{Z}_{2}$ gauge
theory high-precision MC results of ref.~\cite{BCGMP-12} confirmed EST
correction $\lambda^{-3}f_{3}\left(  C\right)  \,$ coming from the boundary
part of the effective action (see sec.~\ref{higher-orders-lambda-section}).

The aim of this work is to test EST for Wilson loops with
\emph{non-rectangular polygonal} contours. In case of large non-rectangular
contours, EST starts to work at rather large sizes of polygons, which imposes
rather severe requirements on computer resources. Therefore concentrating on
the problem of non-rectangular contours one has to sacrifice other degrees of
freedom in lattice tests: the current work deals with

-- the minimal non-trivial case of $D=3$ space-time dimensions,

-- the simplest gauge group $\mathbb{Z}_{2}$ (near its critical point where a
full-fledged continuous quantum field theory is generated).

Another price that one has to pay for working with non-rectangular polygons is
an insufficient accuracy of MC results, which makes difficult testing
higher-order terms of the EST large-size expansion.

\subsection{Structure of the paper}

Sections \ref{section-Wilson-loops-EST} -- \ref{section-Laplace-determinants}
contain a review of well-known facts needed for understanding the main part of
this work. In sec.~\ref{section-Wilson-loops-EST} a brief derivation of
asymptotic EST\ formulas (\ref{W-product-det-product-0}),
(\ref{W-product-det-product-1}) is sketched. Sec.
\ref{z2-gauge-theory-section} describes basic properties of lattice 3D
$\mathbb{Z}_{2}$ gauge theory and the duality of this theory to 3D Ising
model. Sec.~\ref{section-Laplace-determinants} is devoted to determinants of
two-dimensional Laplace operators. The computation of Laplace determinants
needed for the comparison of MC\ results with EST is described in Appendix.

In sec.~\ref{polygons-choice-section} we discuss the optimal choice of
polygons for MC lattice tests of EST. In sec.~\ref{section-lattice-results}
our lattice results are summarized and compared with EST.

Expert readers interested in the immediate comparison of lattice MC results
with EST predictions may jump directly to Table~2 and to the content of sec.
\ref{method-1-section} which provides a simple but incomplete analysis of
results. Sec.~\ref{method-2-section} contains a more thorough analysis of data
by combining the results of this work with the high-precision value for the
string tension $\sigma$ of 3D $\mathbb{Z}_{2}$ gauge theory from ref.
\cite{BCGMP-12}.

\section{Wilson loops in EST}

\setcounter{equation}{0} 

\label{section-Wilson-loops-EST}

\subsection{Basic ideas}

As already mentioned above, the main idea of EST is that in systems where ESF
occurs, low-energy and large-distance properties of the effective string can
be described by an effective string action. The implementation of this idea is
analogous to systems with a spontaneous breakdown of continuous symmetries:

-- Identify low-energy degrees of freedom corresponding to the broken symmetry
(partial breakdown Poincar\'{e} group in case of ESF).

-- Write the most general action containing all terms compatible with the
symmetries of the microscopic gauge theory.

-- Effective Lagrangian is an infinite series of terms ordered according to
the counting of gradients.

-- All terms compatible with the symmetries of the microscopic gauge theory
are allowed (but at a deeper level of the analysis certain constraints may be derived).

-- This effective action allows to compute systematically low-energy and
large-distance expansions for various quantities. In particular, large-size
expansion (\ref{W-lambda-C-general-expansion-0}) can be constructed for Wilson loops.

-- Although the effective action contains an infinite amount of terms, for the
computation of any given order of the large-size expansion only a finite
amount of terms of the effective action is needed.

-- Coefficients accompanying the terms of the effective action are typically
determined by the microscopic theory and cannot be computed in EST. However,
symmetries and general principles of quantum field theory may impose certain
constraints on these coefficients.

-- Loop diagrams of the effective theory may contain ultraviolet\ divergences.
From the physical point of view these divergences must be cut at the boundary
of the applicability of the effective theory. However, technically in
practical computations the formal treatment of these divergences has many
common features with the usual renormalization.

\subsection{One-loop expression for $W\left(  C\right)  $ in EST}

\label{EST-one-loop-section}

The idea that in some limit Wilson loops $W\left(  C\right)  $ can be
approximated by the functional integral over surfaces $\Sigma$ whose boundary
is fixed at Wilson contour~$C$
\begin{equation}
W\left(  C\right)  \rightarrow\mathrm{const}\int_{\partial\Sigma=C}D\Sigma
\exp\left(  -S_{\mathrm{EST}}\left[  \Sigma\right]  \right)  \,.
\label{W-C-EST}
\end{equation}
has a long history \cite{Polyakov-1980}. The detailed implementation of this
idea depends on which asymptotic limit is assumed: large-size
\cite{Luscher:1980fr} or large-$N$ \cite{MM-1981}, \cite{Migdal-1984}. As was
already mentioned, the large-$N$ case is not related to the content of this
work but it is an interesting case, especially when the large-$N$ limit is
combined with supersymmetry and with AdS/CFT correspondence.

The large-size limit allows to arrange the effective string action
$S_{\mathrm{EST}}\left[  \Sigma\right]  $ as an infinite series of terms
according to the counting of gradients. The leading-order part of EST action
is Nambu action
\begin{equation}
S_{\mathrm{EST}}\left(  \Sigma\right)  =\sigma_{0}A\left(  \Sigma\right)
+\ldots\label{S-EST-Nambu-start}
\end{equation}
Here $A\left(  \Sigma\right)  $ is the area of $\Sigma$, $\sigma_{0}$ is the
bare string tension. Generally one should distinguish between bare string
action $\sigma_{0}$ appearing in (\ref{S-EST-Nambu-start}) and the physical
string tension $\sigma$ appearing in area law (\ref{Wilson-area-law}). But in
some regularizations automatically removing power divergences (e.g.
dimensional and $\zeta$ regularizations) the difference between $\sigma_{0}$
and $\sigma$ is not visible in the computation of the first orders of the
large-size expansion.

The case when one approximates $S_{\mathrm{EST}}\left(  \Sigma\right)  $ by
Nambu action $\sigma_{0}A\left(  \Sigma\right)  $ and computes (\ref{W-C-EST})
in one-loop approximation using the steepest-decent method was studied in ref.
\cite{Luscher:1980fr} for smooth contours $C$. Below we briefly sketch the
generalization of results of ref.~\cite{Luscher:1980fr} to the case of flat
\emph{polygonal} contours.

For the computation of the first orders of the large-size expansion of Wilson
loops, the next-to-Nambu corrections in eq.~(\ref{S-EST-Nambu-start}) are
irrelevant so that one has to compute
\begin{equation}
\int_{\partial\Sigma=C}D\Sigma\exp\left[  -\sigma_{0}A\left(  \Sigma\right)
\right]  \,.
\end{equation}
Surface area defining Nambu action can be represented in the form
\begin{equation}
A\left(  \Sigma\right)  =\int_{U}du^{1}du^{2}\sqrt{\det_{a,b}\left(
Q_{ab}\right)  }\,,
\end{equation}
\begin{equation}
Q_{ab}=\sum_{\mu,\nu=1}^{D}\frac{\partial X^{\mu}}{\partial u^{a}
}\frac{\partial X^{\mu}}{\partial u^{b}}\,.
\end{equation}
Nambu action is reparametrization invariant. Therefore in EST\ (like in the
theory of fundamental boson strings) reparametrization invariance must be
treated as a gauge symmetry. In case of flat Wilson contours $C$, it is
convenient to use the planar gauge (sometimes called static gauge):
\begin{equation}
u^{a}=X^{a}\quad\left(  a=1,2\right)  \,,
\end{equation}
\begin{equation}
X^{\perp}=X^{\perp}\left(  X^{1},X^{2}\right)  \quad\left(  \perp=3,4,\ldots
D\right)  \,.
\end{equation}
In this gauge
\begin{equation}
Q_{ab}=\delta_{ab}+\sum_{k=3}^{D}\frac{\partial X^{k}}{\partial X^{a}
}\frac{\partial X^{k}}{\partial X^{b}}\,,
\end{equation}
In quadratic approximation
\begin{equation}
\sqrt{\det_{a,b}\left(  Q_{ab}\right)  }=1+\frac{1}{2}\sum_{a=1}^{2}\sum
_{k=3}^{D}\frac{\partial X^{k}}{\partial X^{a}}\frac{\partial X^{k}}{\partial
X^{a}}+\ldots
\end{equation}
\begin{equation}
\iint_{U}dX^{1}dX^{2}\sqrt{\det_{a,b}\left(  Q_{ab}\right)  }=S\left(
C\right)  +\frac{1}{2}\iint_{U}dX^{1}dX^{2}\sum_{a=1}^{2}\sum_{k=3}
^{D-2}\frac{\partial X^{k}}{\partial X^{a}}\frac{\partial X^{k}}{\partial
X^{a}}\,,
\end{equation}
\begin{equation}
\partial U=C\,.
\end{equation}
The steepest-descent integration yields
\begin{equation}
\int_{\partial\Sigma=C}D\Sigma\exp\left[  -\sigma_{0}A\left(  \Sigma\right)
\right]  =\mathrm{const}\,\left[  \mathrm{Det}_{\mathrm{reg}}\left[
-\Delta\left(  C\right)  \right]  \right]  ^{-\left(  D-2\right)  /2}
\exp\left[  -\sigma_{0}S\left(  C\right)  \right]  \,.
\end{equation}
For polygonal contours Laplace determinant has area, perimeter and cusp
divergences so that the regularized and renormalized Laplace determinants are
connected by the relation
\begin{equation}
\mathrm{Det}_{\mathrm{reg}}\left[  -\Delta\left(  C\right)  \right]
=\exp\left[  a_{2}S\left(  C\right)  +a_{1}L\left(  C\right)  +a_{0}\left(
C\right)  \right]  \mathrm{Det}_{\mathrm{ren}}\left[  -\Delta\left(  C\right)
\right]  \,.
\end{equation}
Explicit expressions for these divergences in proper-time regularization are
given by eq.~(\ref{divergences-proper-time}). In $\zeta$-regularization, power
divergences $a_{1},a_{2}$ vanish and one is left only with the cusp divergence
$a_{0}\left(  C\right)$
\begin{equation}
\zeta\text{-regularization:\quad}a_{1}=0,\quad a_{2}=0\,. \label{zeta-a-1-a-2}
\end{equation}

After the renormalization one arrives at the asymptotic expansion

\begin{align}
&  W\left(  C\right)  \overset{\left|  C\right|  \rightarrow\infty}{=}K\left[
\prod_{\gamma=1}^{M\left(  C\right)  }B\left(  \theta_{\gamma}\right)
\right]  \exp\left[  -\sigma S\left(  C\right)  +\rho L\left(  C\right)
\right] \nonumber\\
&  \times\left[  \mathrm{Det}_{\zeta}\left(  -\Delta\left(  C\right)  \right)
\right]  ^{-(D-2)/2}\left(  1+\ldots\right)  \,.
\label{W-C-asymptotic-EST-general}
\end{align}

We use $\zeta$-renormalization scheme for the renormalized Laplace
determinant. In principle, other renormalization schemes could be also used up
to some subtleties:

1) In the final part of our work we will need explicit expressions for Laplace
determinants on hexagons which will be computed using general expressions of
ref. \cite{AS-93-journal} where Laplace determinants were computed in the $\zeta
$-regularization scheme.

2) $\zeta$-regularization has important properties (\ref{zeta-a-1-a-2}),
(\ref{Laplace-det-zeta-scaling}). They allow to identify parameter $\sigma$
appearing in eq.~(\ref{W-C-asymptotic-EST-general}) with the physical string tension.

In eq.~(\ref{W-C-asymptotic-EST-general}), $\theta_{\gamma}$ are interior
vertex angles of the polygon. $M\left(  C\right)  $ is the number of vertices
of the polygon.

Parameters $K,\rho,\sigma$ and function $B\left(  \theta_{\gamma}\right)  $
appearing on the RHS\ of eq.~(\ref{W-C-asymptotic-EST-general}) are determined
by the dynamics of the microscopic gauge theory and cannot be computed in EST.
For flat polygons $C$, sum rule
\begin{equation}
\sum_{\gamma}\frac{\pi-\theta_{\gamma}}{2\pi}=1 \label{angle-sum-rule}
\end{equation}
allows to minimize the amount of unknown factors (not controlled by EST) by
redefining
\[
\tilde{B}\left(  \theta_{\gamma}\right)  =\left[  B\left(  \theta_{\gamma
}\right)  K^{\left(  \pi-\theta_{\gamma}\right)  /(2\pi)}\right]
\]
so that
\begin{equation}
K\left[  \prod_{i=1}^{M\left(  C\right)  }B\left(  \theta_{\gamma}\right)
\right]  =\prod_{i=1}^{M\left(  C\right)  }\tilde{B}\left(  \theta_{\gamma
}\right)  \,. \label{B-B-tilde-reduction}
\end{equation}
On the other hand, coefficient $K$ provides a natural way to represent the
freedom of normalization of Wilson loops in the microscopic gauge theory.

Combining eq.
(\ref{W-C-asymptotic-EST-general}) with scaling property
(\ref{Laplace-det-zeta-scaling}) of $\mathrm{Det}_{\zeta}\left[
-\Delta\left(  \lambda C\right)  \right]  $ we arrive at area law
(\ref{Wilson-area-law}). This means that parameter $\sigma$ appearing in
eq.~(\ref{W-C-asymptotic-EST-general}) has the meaning of the ``observable''
physical string tension.

Using property (\ref{log-Laplace-det-zeta-scaling}) of Laplace determinant in
$\zeta$-regularization and eq.~(\ref{Kac-dimension}), we can rewrite
(\ref{W-C-asymptotic-EST-general}) in the form
\begin{align}
&  \ln W\left(  \lambda C\right)  \overset{\lambda\rightarrow\infty}{=}
\ln\left(  \prod_{i=1}^{M\left(  C\right)  }\tilde{B}\left(  \theta_{\gamma
}\right)  \right)  -\sigma\lambda^{2}S\left(  C\right)  +\rho\lambda L\left(
C\right) \nonumber\\
&  +\frac{D-2}{2}\left\{  \sum\limits_{\gamma=1}^{M\left(  C\right)  }
\xi(\theta_{\gamma})\ln\lambda-\ln\left[  \mathrm{Det}_{\zeta}\left(
-\Delta\left(  C\right)  \right)  \right]  \right\}  +O\left(  \lambda
^{-2}\right)  \label{log-W-one-loop}
\end{align}
where function $\xi(\theta)$ is given by eq.~(\ref{Kac-dimension}). The status
of the higher-order corrections starting from $O\left(  \lambda^{-2}\right)  $
will be discussed in sec.~\ref{higher-orders-lambda-section}.

Thus we have the following structure of the large-$\lambda$ expansion
corresponding to the first terms of the general expansion
(\ref{W-lambda-C-general-expansion-0}):
\begin{equation}
\ln W\left(  \lambda C\right)  \overset{\lambda\rightarrow\infty}{=}f_{\ln
}\left(  C\right)  \ln\lambda+\lambda^{2}f_{-2}\left(  C\right)  \,+\lambda
f_{-1}\left(  C\right)  \,+f_{0}\left(  C\right)  \,+O\left(  \lambda
^{-2}\right)  \label{W-lambda-C-general-one-loop-expansion}
\end{equation}
where
\begin{equation}
f_{\ln}\left(  C\right)  =\frac{D-2}{2}\sum\limits_{\gamma=1}^{M\left(
C\right)  }\xi(\theta_{\gamma})\,,\label{f-ln}%
\end{equation}%
\begin{equation}
f_{-2}\left(  C\right)  =-\sigma S\left(  C\right)  \,, \label{f-minus-2-S}
\end{equation}
\begin{equation}
f_{-1}\left(  C\right)  =\rho L\left(  C\right)  \,, \label{f-minus-1-L}
\end{equation}
\begin{equation}
f_{0}\left(  C\right)  =\ln\left(  \prod_{i=1}^{M\left(  C\right)  }\tilde
{B}\left(  \theta_{\gamma}\right)  \right)  -\frac{D-2}{2}\ln\left[
\mathrm{Det}_{\zeta}\left(  -\Delta\left(  C\right)  \right)  \right]  \,.
\end{equation}

\subsection{Derivation of the asymptotic formula for a ``balanced'' product of
Wilson loops}
\label{derivation-asymptotic-formula-section}

Now we turn to the derivation of eq.~(\ref{W-product-det-product-1}).
As mentioned in the
comments to eq.~(\ref{W-product-det-product-1}), we assume that we have
several polygonal contours $C_{i}$ and a set of numbers $m_{i}$ obeying
conditions (\ref{balance-perimeter-0}). We also assume that vertex angles
$\theta_{\gamma i}$ of polygons $C_{i}$ belong to a common set of angles
$\left\{  \theta_{a}\right\}  $, angle $\theta_{a}$ appearing $p_{ia}$ times
in $C_{i}$ with constraint (\ref{balance-angles-0}). Then according to
eq.~(\ref{f-ln})
\begin{equation}
\sum_{i}m_{i}f_{\ln}\left(  C_{i}\right)  =\frac{D-2}{2}\sum_{i}m_{i}
\sum\limits_{\gamma=1}^{M\left(  C\right)  }\xi(\theta_{\gamma i})\,.
\end{equation}
Obviously
\begin{equation}
\sum\limits_{\gamma=1}^{M\left(  C\right)  }\xi(\theta_{\gamma i}
)=\sum\limits_{a}p_{ia}\xi(\theta_{a})
\end{equation}
so that
\begin{equation}
\sum_{i}m_{i}f_{\ln}\left(  C_{i}\right)  =\frac{D-2}{2}\sum\limits_{a=1}
^{M}\xi(\theta_{a})\left(  \sum_{i}m_{i}p_{ia}\right)  \,.
\end{equation}
Then constraint (\ref{balance-angles-0}) leads to
\begin{equation}
\sum_{i}m_{i}f_{\ln}\left(  C_{i}\right)  =0\,.\label{sum-f-ln-0}
\end{equation}
Similarly we have
\begin{equation}
\sum_{i}m_{i}\ln\left(  \prod_{\gamma=1}^{M\left(  C_{i}\right)  }\tilde
{B}\left(  \theta_{\gamma i}\right)  \right)  =\sum\limits_{a=1}^{M}\left(
\ln\tilde{B}\left(  \theta_{a}\right)  \right)  \left(  \sum_{i}m_{i}%
p_{ia}\right)  =0\,.\label{sum-ln-B-0}
\end{equation}
Now we find from eqs. (\ref{log-W-one-loop}), (\ref{sum-f-ln-0}),
(\ref{sum-ln-B-0})
\begin{align}
&  \sum_{i}m_{i}\ln W\left(  \lambda C_{i}\right)  \overset{\lambda
\rightarrow\infty}{=}-\sigma\lambda^{2}\sum_{i}m_{i}S\left(  C_{i}\right)
\nonumber\\
&  -\frac{D-2}{2}\sum_{i}m_{i}\ln\left[  \mathrm{Det}_{\zeta}\left(
-\Delta\left(  C_{i}\right)  \right)  \right]  +O\left(  \lambda^{-2}\right)
\,.\label{W-product-det-product-1-pre}
\end{align}
This proves asymptotic formula (\ref{W-product-det-product-1}) under
assumptions (\ref{balance-perimeter-0}), (\ref{balance-angles-0}). If one adds
area constraint (\ref{balance-area-0}) then eq.~(\ref{W-product-det-product-1})
simplifies to eq.~(\ref{W-product-det-product-0}).

\subsection{Dependent constraint}

\label{dependent-constraint-section}

Note that our derivation of eq.~(\ref{W-product-det-product-1-pre}) for
polygonal contours $C_{i}$ did not use constraint (\ref{balance-vertices-0}).
Technically we could avoid the involvement of eq.~(\ref{balance-vertices-0})
by using reduction (\ref{B-B-tilde-reduction}) but the real reason is that eq.
(\ref{balance-vertices-0}) is a trivial consequence of constraint
(\ref{balance-angles-0}) and sum rule (\ref{angle-sum-rule}). Indeed, eq.
(\ref{angle-sum-rule}) can be rewritten in the form
\begin{equation}
1=\sum_{a}p_{ia}\frac{\pi-\theta_{a}}{2\pi}
\end{equation}
which leads to
\begin{equation}
\sum_{i=1}^{n}m_{i}=\sum_{i=1}^{n}m_{i}\sum_{a}p_{ia}\frac{\pi-\theta_{a}
}{2\pi}=\sum_{a}\frac{\pi-\theta_{a}}{2\pi}\sum_{i=1}^{n}m_{i}p_{ia}\,.
\end{equation}

\subsection{Higher orders of the large-size expansion}

\label{higher-orders-lambda-section}

One-loop result (\ref{W-lambda-C-general-one-loop-expansion}) can be
generalized to higher orders of the large-$\lambda$ expansion according to eq.
(\ref{W-lambda-C-general-expansion-0}). A careful derivation of expansion
(\ref{W-lambda-C-general-expansion-0}) requires much work:

1) analysis of constraints on terms of the full EST action
\cite{Luscher:2004ib}-\cite{BCGMP-12},

2) computation of diagrams generated by this effective action \cite{BCGMP-12},
\cite{Filk-preprint}-\cite{Billo:2011fd}.

As was already mentioned in eq.~(\ref{f-1-zero}), $f_{1}\left(  C\right)  =0$.
The reason is that EST\ action contains two parts:

-- surface action represented by a two-dimensional integral over the region
bounded contour $C$ (including Nambu action),

-- boundary action represented by a one-dimensional integral over contour $C$
\cite{Luscher:2004ib}, \cite{Aharony:2010cx}, \cite{BCGMP-12}

\noindent The surface action generates terms $f_{n}\left(  C\right)  $
with even $n$. The boundary action also generates terms with odd $n$ but only
starting from $f_{3}\left(  C\right)  $. Hence neither surface nor boundary
actions can generate $f_{1}\left(  C\right)  $.

One should be alerted about the traditional but somewhat confusing naming
scheme for terms $f_{n}\left(  C\right)  $ of expansion
(\ref{W-lambda-C-general-expansion-0}). Terms $f_{2n}\left(  C\right)  $ are
often called $\left(  n+1\right)  $-loop corrections implying that these terms
come from $\left(  n+1\right)  $-loop diagrams with vertices generated by the
surface terms of the effective action. This nomenclature ignores contributions
from the boundary action, in particular terms $f_{n}\left(  C\right)  $ with
odd $n$.

Correction $f_{2}\left(  C\right)  $ is in principle computable in EST but
currently it is known only for rectangular Wilson contours $C$. This
correction was computed for rectangular Wilson loops and for the correlation
function of two Polyakov loops in refs.~\cite{Filk-preprint}, \cite{Dietz-83}. The
result of refs.~\cite{Filk-preprint}, \cite{Dietz-83} for Polyakov loops was
confirmed by a computation in another regularization \cite{Luscher:2004ib}.
However, the result of refs.~\cite{Filk-preprint}, \cite{Dietz-83} for Wilson loops
has an arithmetic error which was corrected in refs. \cite{BCVG-2010},
\cite{Billo:2011fd}.

Correction $f_{3}\left(  C\right)  $ comes from the boundary part of the
effective action. In case of the correlation function of two Polyakov lines
this correction was computed in ref.~\cite{Aharony:2010cx}. In case of
rectangular Wilson loops the $f_{3}\left(  C\right)  $ term for was computed
in EST and tested by lattice MC (in 3D $\mathbb{Z}_{2}$ gauge theory) in ref.
\cite{BCGMP-12}.

\section{Lattice $\mathbb{Z}_{2}$ gauge theory}

\label{z2-gauge-theory-section}

\setcounter{equation}{0} 

\subsection{$\mathbb{Z}_{2}$ gauge theory}

High-precision tests of EST are often performed using $\mathbb{Z}_{2}$ gauge
theory (sometimes called gauge Ising model) in $D=3$ Euclidean space-time
dimensions \cite{BCGMP-12}, \cite{Caselle:2005xy}, \cite{Caselle:2006dv}.
Discrete group $\mathbb{Z}_{2}$ contains two elements $\pm1$. $\mathbb{Z}_{2}$
gauge theory is defined by the standard Wilson lattice action with link
variables $U_{l}=\pm1$. In particular, the partition function $Z_{\mathbb{Z}_{2}}$
of $\mathbb{Z}_{2}$ gauge theory is
\begin{equation}
Z_{\mathbb{Z}_{2}}\left(  \beta_{\mathbb{Z}_{2}}\right)  =\sum_{\left\{
U_{l}\right\}  }\exp\left[  \sum_{P}\beta_{\mathbb{Z}_{2}}\Pi\left(  P\right)
\right]  \,.\label{Z-Z-2-def}
\end{equation}
On the RHS, the external sum  runs over all lattice link configurations
$\left\{  U_{l}\right\}  $ with $U_{l}=\pm1$. $\beta_{\mathbb{Z}_{2}}$ is
lattice coupling constant, the sum in the exponent runs over all non-oriented
plaquettes $P$, and $\Pi\left(  P\right)  =\pm1$ is the product of link
variables $U_{l}=\pm1$ associated with four links $P_{i}$ of plaquette $P$
\begin{equation}
\Pi\left(  P\right)  =\prod_{l\in P}U_{l}\,.
\end{equation}

There are several reasons why 3D $\mathbb{Z}_{2}$ gauge theory is frequently
used in lattice MC\ simulations:

1) Many predictions of EST are universal, i.e. independent of the gauge group
[see e.g. eq.~(\ref{W-product-det-product-0})]. Therefore in high-precision MC
lattice tests of EST it is natural to choose the simplest but still nontrivial
confining gauge theory. 3D $\mathbb{Z}_{2}$ gauge theory is the best candidate
for this role.

2) $D=3$ is the minimal space-time dimension where area law and EST are
implemented nontrivially. At $D=2$, non-Abelian gauge theories are exactly
solvable, EST corrections to the area law are proportional to $D-2$ (number of
the transverse degrees of freedom) and vanish at $D=2$.

3) Although $\mathbb{Z}_{2}$ group is Abelian, the $\mathbb{Z}_{2}$ lattice
gauge theory has a confining phase.

4) 3D $\mathbb{Z}_{2}$ gauge theory is dual to 3D Ising model. This duality is
crucial for understanding the confining properties and the continuous limit of
3D $\mathbb{Z}_{2}$ gauge theory. $\mathbb{Z}_{2}$-Ising duality also plays an
important role in high-efficiency algorithms of lattice MC\ simulations.

\subsection{Duality of 3D Ising model and lattice $\mathbb{Z}_{2}$ gauge theory}

\label{Ising-Z2-duality-section}

\subsubsection{Preliminary comments}

This section contains a brief review of 3D $\mathbb{Z}_{2}$-Ising duality.
Before turning to technical details it makes sense to describe the main
consequences of this duality.

1) 3D lattice $\mathbb{Z}_{2}$ gauge theory has confinement and deconfinement
phases separated by a critical point. $\mathbb{Z}_{2}$-Ising duality maps the
confinement phase of $\mathbb{Z}_{2}$ theory to the spontaneously broken phase
of Ising model whereas the $\mathbb{Z}_{2}$ deconfinement phase is mapped to
the Ising non-broken phase.

2) Due to $\mathbb{Z}_{2}$-Ising duality the critical point of $\mathbb{Z}_{2}$
theory belongs to the same universality class as the critical point of
3D\ Ising model. This critical point can be used for the construction of
various continuous field theories. If the lattice theory is taken exactly at
the critical point then its correlation functions have a power asymptotic
behavior at large distances, which is controlled by the conformal field theory
corresponding to this universality class.

However, if one slightly deviates from the critical point towards the
confinement phase of $\mathbb{Z}_{2}$ theory (or towards the broken phase of
Ising model)  and combines the approach to the critical point with an
appropriate rescaling of distances and momenta then one arrives at a quite
physical full-fledged continuous quantum field theory with \emph{massive}
particles. This massive continuous quantum field theory exists in two
essentially equivalent $Z_{2}$-based and Ising-based versions. In
$Z_{2}$-representation, massive particles can be interpreted as ``glueballs'' whose
spectrum can be measured by lattice MC \cite{ACCH-1997}. Wilson loops of
$Z_{2}$-based theory obey area law (\ref{Wilson-area-law}).

3) $\mathbb{Z}_{2}$-Ising duality allows to interpret the confinement of
$\mathbb{Z}_{2}$ gauge theory in terms of Ising model. In particular, the
string tension of confining strings in $\mathbb{Z}_{2}$ theory is equal to the
surface tension between two different spontaneously broken phases of Ising
model. The $\mathbb{Z}_{2}$-Ising duality provides a correspondence between
correlation functions of the two dual theories. $\mathbb{Z}_{2}$ Wilson loops
can be expressed via multi-spin correlation functions of Ising model.

4) One can also profit from the $\mathbb{Z}_{2}$-Ising duality in lattice MC
simulations. The computation of $\mathbb{Z}_{2}$ Wilson loops can be performed
directly in Ising model. Numerical simulations in 3D Ising model have a long
history. Powerful algorithms have been developed. In particular, for the
computation of $\mathbb{Z}_{2}$ Wilson loops directly in the dual Ising model
an extremely efficient hierarchical MC (HMC) algorithm was suggested and successfully
applied (see sec.~\ref{hierarchical-algorith-section}). One of the features of
this method is that it allows to compute a ratio of two Wilson loops directly,
without computing the two Wilson loops separately. Large Wilson loops make
many problems in\ lattice MC computations. The possibility to avoid their
separate computation and to work directly with ratios of Wilson loops allows
to compute ratios of large Wilson loops in 3D $\mathbb{Z}_{2}$ gauge theory
with an unprecedented accuracy.

\subsubsection{Ising model}

The partition function of Ising model is

\begin{equation}
Z_{I}\left(  \beta_{I}\right)  =\sum_{\left\{  s_{k}\right\}  }\exp\left[
\beta_{I}\sum_{\{i,j\}\text{: next neighbors}}s_{i}s_{j}\right]  \,.
\end{equation}
The external sum on the RHS runs over all lattice site configurations
$\left\{  s_{k}\right\}  $ with $s_{k}=\pm1$. The sum in the exponent runs
over links represented by non-ordered pairs$\{i,j\}$ of next-neighbor sites
$i,j$. $\beta_{I}$ is lattice coupling constant (inverse temperature in the
classical interpretation of Ising model).

Ising model can be considered for arbitrary dimension $D$. As is well known, the
2D Ising model is exactly solvable. It is remarkable that the critical point
of 2D Ising model was first computed using the duality by Kramers and Wannier
\cite{KW-1941} before Onsager \cite{Onsager-1944} found the exact solution of
the model (computing the free energy at arbitrary temperatures). The duality
transformation maps high temperatures to low temperatures. In terms of inverse
temperature $\beta$ duality transformation $\beta_{I,2D}\rightarrow
\beta_{I,2D}^{\prime}$ is described by the relation
\begin{equation}
\sinh\left(  2\beta_{I,2D}^{\prime}\right)  \,\sinh\left(  2\beta
_{I,2D}\right)  =1\,.\label{duality-Ising-2D}
\end{equation}
This transformation leaves the critical point of 2D Ising model intact:
\begin{equation}
\beta_{I,2D}^{\prime}=\beta_{I,2D}=\beta_{I,2D}^{\mathrm{crit}}
\end{equation}
so that eq.~(\ref{duality-Ising-2D}) leads to the following value for the
critical point:
\begin{equation}
\beta_{I,2D}^{\mathrm{crit}}=\frac{1}{2}\ln\left(  1+\sqrt{2}\right)  \,.
\end{equation}

The case of 3D Ising model is different: in spite of much invested effort no
exact solution was found. But it is possible to extend the concept of duality
to $D>2$ \cite{Wegner-1971} and to show that 3D Ising model is dual to lattice
$\mathbb{Z}_{2}$ gauge theory \cite{BDI-1975} (see also review
\cite{Savit-1980} and references therein).

\subsubsection{Duality of 3D Ising model and $\mathbb{Z}_{2}$ gauge theory}

3D $\mathbb{Z}_{2}$ lattice gauge theory with coupling constant $\beta
_{\mathbb{Z}_{2}}$ (\ref{Z-Z-2-def}) is dual to Ising model with coupling
$\beta_{I}$ if constants $\beta_{\mathbb{Z}_{2}}$ and $\beta_{I}$ are
connected by relation \cite{BDI-1975}
\begin{equation}
\sinh\left(  2\beta_{I}\right)  \sinh\left(  2\beta_{\mathbb{Z}_{2}}\right)
=1,\label{duality-beta-1}
\end{equation}
which formally coincides with duality relation (\ref{duality-Ising-2D}) for
the self-dual 2D Ising model. But now we deal with a duality connecting two
different models.

$\mathbb{Z}_{2}$-Ising duality relations can be derived for many quantities
(free energy, correlation functions), see refs. \cite{BDI-1975},
\cite{Savit-1980}. For our work we need only the expression for Wilson loops
of $\mathbb{Z}_{2}$ gauge theory via multi-spin correlation functions of Ising
model \cite{BDI-1975}, \cite{Savit-1980}, \cite{Caselle:1995fh},
\cite{Caselle:1996ii}. The Ising-model representation for the $\mathbb{Z}_{2}$
Wilson loop $W\left(  C\right)  $ can be constructed in several steps:

1) Choose some lattice surface $S$ bounded by contour $C$ of the Wilson loop
(the choice of $S$ is not important but for flat contours $C$ it is natural to
take the flat region bounded by $C$).

2) Select all next-neighbor Ising spin links crossing surface $S$
($\mathbb{Z}_{2}$ and Ising models live on dual lattices, i.e. links of the
Ising model are in one-to-one correspondence to plaquettes of $\mathbb{Z}_{2}$
theory crossed by Ising links).

3) Change the sign of Ising coupling $\beta_{I}$ for the selected links in the
Gibbs distribution of Ising model.

4) Compute partition function $Z_{S}$ corresponding to the Ising model with
the above change of the spin couplings.

5) Normalize $Z_{S}$ with respect to the usual Ising partition function $Z$.
The result $Z_{S}/Z$ is the Ising dual representation for Wilson loop
$W\left(  C\right)$ of $\mathbb{Z}_{2}$ theory:

\begin{equation}
W\left(  C\right)  =\frac{Z_{S}}{Z}\,.\label{W-dual-BCP}
\end{equation}
In case of two Wilson loops eq.~(\ref{W-dual-BCP}) leads to
\begin{equation}
\frac{W\left(  C_{1}\right)  }{W\left(  C_{2}\right)  }=\frac{Z_{S_{1}}
}{Z_{S_{2}}}\,.\label{W-ratio-Z-ratio}
\end{equation}
If $S_{2}\subset S_{1}$ (i.e. contour $C_{1}$ encloses contour $C_{2}$) then
eq.~(\ref{W-ratio-Z-ratio}) can be rewritten as
\begin{equation}
\frac{W\left(  C_{1}\right)  }{W\left(  C_{2}\right)  }=\left\langle
\exp\left[  -2\beta_{I}\sum_{\left(  i,j\right)  \in D\left(  S_{1}\backslash
S_{2}\right)  }s_{i}s_{j}\right]  \right\rangle _{S_{2}}
\,.\label{W-ratio-S2-average}
\end{equation}
Here $S_{1}\backslash S_{2}$ is the complement of $S_{2}$ in $S_{1}$.
$D\left(  S_{1}\backslash S_{2}\right)  $ is the set of Ising links dual to
$\mathbb{Z}_{2}$-lattice surface $S_{1}\backslash S_{2}$. The sum in the
exponent runs over pairs $(i,j)$ of Ising neighbor sites which are dual to
surface $S_{1}\backslash S_{2}$. Notation $\left\langle \ldots\right\rangle
_{S_{2}}$ stands for averaging with the statistical weight corresponding to
partition function $Z_{S_{2}}$.

\subsubsection{Critical point}

The critical point of 3D Ising model is known with a high precision
\cite{DB-2003}:
\begin{equation}
\beta_{I}^{\mathrm{crit}}=0.22165455(3)\,. \label{beta-Ising-critical}
\end{equation}
Using duality relation (\ref{duality-beta-1}), one finds the critical point of
3D $\mathbb{Z}_{2}$ gauge theory:
\begin{equation}
\beta_{\mathbb{Z}_{2}}^{\mathrm{crit}}=0.76141346(7)\,.
\label{beta-Z2-critical}
\end{equation}

\subsubsection{Hierarchical MC algorithm for ratios of large Wilson loops}

\label{hierarchical-algorith-section}

Lattice tests of EST must deal with Wilson loops of large size. A
straightforward MC computation of large Wilson loops requires computer time
increasing exponentially with growth of the area of the Wilson loop. The
problem of this exponential growth is partly solved by L\"{u}scher-Weisz
algorithm \cite{LW-01}, \cite{Luscher:2002qv} which can be used for any gauge
group. In case of lattice $\mathbb{Z}_{2}$ gauge theory another high-efficiency
algorithm was suggested in ref. \cite{Caselle:2002ah}. In applications to
Wilson loops this algorithm is well described in ref. \cite{Billo:2011fd}.

This algorithm is based on several ideas.

1) If one is interested in a MC computation of a ratio of two large Wilson
loops $W\left(  C\right)  /W\left(  C^{\prime}\right)  $ then one should
compute this ratio directly without splitting the problem in separate MC
computations of $W\left(  C\right)  $ and $W\left(  C^{\prime}\right)  $.

2) One should profit from $\mathbb{Z}_{2}$-Ising duality relation
(\ref{W-ratio-S2-average}) and perform the MC\ simulation in the dual Ising model.

3) Instead of the direct computation of $W\left(  C\right)  /W\left(
C^{\prime}\right)  $ in Ising model using eq.~(\ref{W-ratio-S2-average}),
first rearrange this ratio
\begin{align}
\frac{W\left(  C\right)  }{W\left(  C^{\prime}\right)  } &  =\frac{W\left(
C_{0}\right)  }{W\left(  C_{1}\right)  }\frac{W\left(  C_{1}\right)
}{W\left(  C_{2}\right)  }\ldots\frac{W\left(  C_{n-1}\right)  }{W\left(
C_{n}\right)  }\frac{W\left(  C_{n}\right)  }{W\left(  C_{n+1}\right)  }\,,\\
C_{0} &  =C,\quad C_{n+1}=C^{\prime}\,.
\end{align}
Auxiliary contours $C_{1},\ldots C_{n}$ must be chosen so that for all $i$
contours $C_{i}$ and $C_{i+1}$ are rather close and ratios $W\left(
C_{i}\right)  /W\left(  C_{i+1}\right)  $ do not deviate too much from 1 (i.e.
one should avoid exponentially large or exponentially small ratios $W\left(
C_{i}\right)  /W\left(  C_{i+1}\right)  $). In this case each auxiliary ratio
$W\left(  C_{i}\right)  /W\left(  C_{i+1}\right)  $ can be computed without
problems by MC\ simulations in Ising model using eq.~(\ref{W-ratio-S2-average}).
In short, hierarchical MC (HMC) algorithm may be written in the form (in
case $S_{m+1}\subset S_{m}$)
\begin{equation}
\left[  \frac{W\left(  C\right)  }{W\left(  C^{\prime}\right)  }\right]
^{\mathrm{HMC}}=\prod_{m=0}^{n}\left\langle \exp\left[  -2\beta_{I}
\sum_{\left(  i,j\right)  \in D\left(  S_{m}\backslash S_{m+1}\right)  }
s_{i}s_{j}\right]  \right\rangle _{S_{m+1}}^{\mathrm{MC}}\,.
\end{equation}

\section{2D Laplace determinants}

\setcounter{equation}{0} 

\label{section-Laplace-determinants}

\subsection{Ultraviolet divergences and renormalization}

In order to use EST asymptotic formulas (\ref{W-product-det-product-0}),
(\ref{W-product-det-product-1}) for large Wilson loops, we need Laplace
determinants. We use notation $\Delta\left(  C\right)  $ for Laplace operator
defined in a flat region bounded by contour $C$ with Dirichlet boundary condition.

Ultraviolet divergences of 2D Laplace determinants are controlled by the
heat-kernel expansion \cite{Luscher:1980fr}, \cite{Pleijel-1954}-\cite{Guilkey-1974}:
\begin{equation}
\mathrm{Tr}\,e^{t\Delta\left(  C\right)  }=\frac{1}{4\pi t}S(C)-\frac{1}
{8\sqrt{\pi t}}L(C)-\frac{1}{2}\delta\left(  C\right)  +O\left(
t^{1/2}\right)  \,.
\end{equation}
For smooth contours $C$ the $t^{0}$ term of this expansion is given by
\begin{equation}
\delta\left(  C\right)  =-\frac{1}{3}.
\end{equation}
For polygonal contours $C$
\begin{equation}
\delta\left(  C\right)  =-\sum\limits_{\gamma=1}^{M\left(  C\right)  }
\xi(\theta_{\gamma})\,,\label{Kac-dimension}
\end{equation}
Here $\theta_{\gamma}$ are interior polygon angles and
\begin{equation}
\xi(\theta)=\frac{\pi^{2}-\theta^{2}}{12\pi\theta}\,.\label{Gamma-cusp-string}
\end{equation}
For cusp function $\xi(\theta)$, an integral representation was found by M.
Kac in ref. \cite{Kac-1966}. Simplified expression (\ref{Gamma-cusp-string})
for $\xi(\theta)$ was obtained by D. B. Ray (the derivation is described in
ref. \cite{McKean-Singer-67}).

In the proper-time regularization
\begin{equation}
\ln\mathrm{Det}_{\tau}\left[  -\Delta\left(  \lambda C\right)  \right]
=-\int_{\tau}^{\infty}\frac{dt}{t}\mathrm{Tr}\,e^{t\Delta\left(  C\right)  }
\end{equation}
at $\tau\rightarrow0$ one has divergences
\begin{equation}
\ln\mathrm{Det}_{\tau}\left[  -\Delta\left(  \lambda C\right)  \right]
\overset{\tau\rightarrow9}{=}-\frac{1}{4\pi\tau}S(C)+\frac{1}{4\sqrt{\pi\tau}
}L(C)-\frac{1}{2}\delta\left(  C\right)  \ln\tau+O\left(  \tau^{0}\right)\,.
\label{divergences-proper-time}
\end{equation}

\thinspace In $\zeta$-regularization method, one first defines the
regularizing $\zeta$-function at $\mathrm{Re}\,s>2$
\begin{equation}
Z_{C}(s)=\mathrm{Sp}\,\left[  (-\Delta\left(  C\right)  \right]
^{-s}=\frac{1}{\Gamma(s)}\int_{0}^{\infty}dt\,t^{s-1}\mathrm{Tr\,}
e^{t\Delta\left(  C\right)  }\,.\label{Z-C-def}
\end{equation}
Then one performs the analytical continuation to $s=0$ ($Z_{C}(s)$ is regular
at $s=0$) and computes the derivative
\begin{equation}
Z_{C}^{\prime}(s)\equiv\frac{d}{ds}Z_{C}(s)\,.
\end{equation}
The determinant in the $\zeta$-regularization is
\begin{equation}
\mathrm{Det}_{\zeta}\left[  -\Delta\left(  C\right)  \right]  =\exp\left[
-Z_{C}^{\prime}(0)\right]  .\label{det-Z-prime}
\end{equation}
Laplace determinant in $\zeta$-regularization has properties

\begin{equation}
\ln\mathrm{Det}_{\zeta}\left[  -\Delta\left(  \lambda C\right)  \right]
=\delta\left(  C\right)  \ln\lambda+\ln\mathrm{Det}_{\zeta}\left[
-\Delta\left(  C\right)  \right]  \,, \label{log-Laplace-det-zeta-scaling}
\end{equation}

\begin{equation}
\mathrm{Det}_{\zeta}\left[  -\Delta\left(  \lambda C\right)  \right]
=\lambda^{\delta\left(  C\right)  }\mathrm{Det}_{\zeta}\left[  -\Delta\left(
C\right)  \right]  \,. \label{Laplace-det-zeta-scaling}
\end{equation}

Using eqs. (\ref{W-C-asymptotic-EST-general}), (\ref{Kac-dimension}),
(\ref{Gamma-cusp-string}), (\ref{Laplace-det-zeta-scaling}), we obtain
coefficient $f_{\ln}\left(  C\right)  $ appearing in large-size expansion
(\ref{W-lambda-C-general-one-loop-expansion})
\begin{equation}
f_{\ln}\left(  C\right)  =-\frac{D-2}{2}\delta\left(  C\right)  =\left(
D-2\right)  \sum\limits_{\gamma=1}^{M\left(  C\right)  }\frac{\pi^{2}
-\theta_{\gamma}^{2}}{24\pi\theta_{\gamma}}\,. \label{f-minus-one-res}
\end{equation}

\subsection{Computation of 2D Laplace determinants}

As is well known, Laplace determinants in 2D regions can be computed using the
conformal anomaly \cite{Polyakov-1981}-\cite{Alvarez-1983}. In order to
compute Laplace determinant $\mathrm{Det}\left(  -\Delta\left(  C\right)
\right)  $ for the region bounded by contour $C$, one has to construct a
conformal mapping of this region to some standard region (semiplane or
circle). In case of polygons this conformal mapping is given by
Schwarz-Christoffel (SC)\ transformation. Thus combining the conformal anomaly and
SC transformation, one can compute Laplace determinants for arbitrary polygons.
However, on this way one must solve two problems:

-- the anomaly-based representation for Laplace determinants has an integral form
and this integral must be computed,

-- this integral representation has cusp divergences which must be renormalized.

These problems were successfully solved by E. Aurell and P. Salomonson in ref.
\cite{AS-93-journal} where 2D Laplace determinant with Dirichlet boundary
condition for an arbitrary polygon was expressed via parameters of SC mapping.
In Appendix we apply the general results of ref. \cite{AS-93-journal} to the
computation of Laplace determinants $\mathrm{Det}_{\zeta}\left[
-\Delta\left(  H_{n,m}\right)  \right]  $ for hexagon contours $H_{n,m}$
(\ref{H-n-m-geometry}) which are needed for the EST analysis of our
MC\ results for Wilson loops. The final numerical results for $\mathrm{Det}
_{\zeta}\left[  -\Delta\left(  H_{n,m}\right)  \right]  $ are listed in Table
4 of Appendix.

\section{Choice of polygons for lattice tests EST}

\setcounter{equation}{0} 

\label{polygons-choice-section}

\subsection{Hexagons $H_{n,m}$}

Using EST, one can construct the large-size expansion for Wilson loops with
any polygonal contours. But when it comes to lattice tests of EST, the choice
of polygons is constrained by several factors. The cubic lattice allows for
polygons with angles $\pi/2$ and $3\pi/2$ only. EST has been well tested for
rectangular Wilson loops. If one goes beyond rectangles then the simplest
polygon with all vertex angles being $\pi/2$ or $3\pi/2$ is a non-convex
hexagon containing one vertex with angle $3\pi/2$ and five vertices with
angles $\pi/2$.

Any polygon can be described by the cyclic sequence of side its lengths $l_{i}$
and interior vertex angles $\alpha_{i,i+1}$ between sides with lengths $l_{i}$
and $l_{i+1}$
\begin{equation}
P\left(  l_{1},\alpha_{1,2},l_{2},\alpha_{2,3},\ldots,l_{n},\alpha
_{n,1}\right)  \,. \label{P-general}
\end{equation}

\begin{figure}[h]
\includegraphics[width=8.0cm]{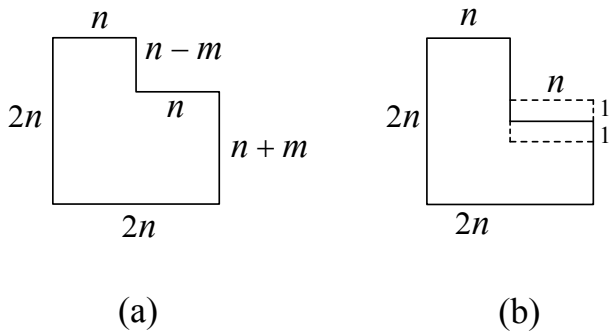} \caption{(a) Hexagon $H_{n,m}$.
(b) Hexagon $H_{n,0}$ (solid) and its deformations (dashed) to $H_{n,1}$ and
$H_{n,-1}$}
\end{figure}

In this paper we work with hexagons (Fig. 1a)
\begin{equation}
H_{n,m}=P\left(  2n,\frac{\pi}{2},2n,\frac{\pi}{2},n,\frac{\pi}{2}
,n-m,\frac{3\pi}{2},n,\frac{\pi}{2},n+m,\frac{\pi}{2}\right)  \,.
\label{H-n-m-geometry}
\end{equation}
These polygons have perimeter
\begin{equation}
L\left(  H_{n,m}\right)  =8n
\end{equation}
and area
\begin{equation}
S\left(  H_{n,m}\right)  =\left(  3n+m\right)  n\,.
\end{equation}

For these hexagons one can define ratios of Wilson loops
\begin{equation}
q_{n,m}^{\pm}=\left[  \frac{W\left(  H_{n,0}\right)  }{W\left(  H_{n,\pm
m}\right)  }\right]  ^{\pm1}\,, \label{q-n-m-def}
\end{equation}
\begin{equation}
r_{n,m}=\frac{W\left(  H_{n,-m}\right)  W\left(  H_{n,m}\right)  }{\left[
W\left(  H_{n,0}\right)  \right]  ^{2}}=\frac{q_{n,m}^{-}}{q_{n,m}^{+}}\,.
\label{r-n-m-def}
\end{equation}
Ratios $q_{n,m}^{\pm}$, $r_{n,m}$ are special cases of the general product
\begin{equation}
\prod_{i=1}^{n}\left[  W\left(  C_{i}\right)  \right]  ^{m_{i}}
\end{equation}
obeying conditions of perimeter and angle balance conditions
(\ref{balance-perimeter-0}), (\ref{balance-angles-0}). In addition, ratios
$r_{n,m}$ obey area balance condition (\ref{balance-area-0}).

In the large-size uniform rescaling limit according to
(\ref{W-product-det-product-1})
\begin{equation}
q_{jn,jm}^{\pm}\overset{j\rightarrow\infty}{=}e^{\sigma j^{2}mn}\left[
\frac{\mathrm{Det}_{\zeta}\left[  -\Delta\left(  H_{n,\pm m}\right)  \right]
}{\mathrm{Det}_{\zeta}\left[  -\Delta\left(  H_{n,0}\right)  \right]
}\right]  ^{\pm(D-2)/2}\left[  1+O\left(  j^{-2}\right)  \right]  \,,
\label{q-uniform-limit}
\end{equation}
\begin{equation}
r_{jn,jm}\overset{j\rightarrow\infty}{=}\left\{  \frac{\mathrm{Det}_{\zeta
}\left[  -\Delta\left(  H_{n,0}\right)  \right]  }{\sqrt{\mathrm{Det}_{\zeta
}\left[  -\Delta\left(  H_{n,m}\right)  \right]  \mathrm{Det}_{\zeta}\left[
-\Delta\left(  H_{n,-m}\right)  \right]  }}\right\}  ^{D-2}\left[  1+O\left(
j^{-2}\right)  \right]  \,. \label{r-uniform-limit}
\end{equation}
In case of $\mathbb{Z}_{2}$ gauge theory, the lattice computation based on
HMC algorithm (see sec.~\ref{hierarchical-algorith-section}) assumes
an independent computation of $q_{n,m}^{-}$ and $q_{n,m}^{+}$. Finally
$r_{n,m}$ can be expressed via $q_{n,m}^{\pm}$ using eq.~(\ref{r-n-m-def}).

In the case of hexagons $H_{n,m}$ (\ref{H-n-m-geometry}) we find from eqs.
(\ref{Kac-dimension}), (\ref{Gamma-cusp-string})
\begin{equation}
\delta\left(  H_{n,m}\right)  =-\frac{5}{9}\,.
\end{equation}
Combining this with (\ref{log-Laplace-det-zeta-scaling}), we obtain
\begin{equation}
\ln\mathrm{Det}_{\zeta}\left[  -\Delta\left(  H_{jn,jm}\right)  \right]
=-\frac{5}{9}\ln j+\ln\mathrm{Det}_{\zeta}\left[  -\Delta\left(
H_{n,m}\right)  \right]  \,. \label{log-det-hexagon-scaling}
\end{equation}
Setting $m=0$, we find
\begin{equation}
\ln\mathrm{Det}_{\zeta}\left[  -\Delta\left(  H_{n,0}\right)  \right]
=\ln\mathrm{Det}_{\zeta}\left[  -\Delta\left(  H_{1,0}\right)  \right]
-\frac{5}{9}\ln n\,. \label{log-det-H-n-n-via-log-det-H-1-1}
\end{equation}

\subsection{Modified large-size limit}

\label{subsection-modified-large-size-limit}

From the theoretical point of view, the uniform rescaling limit
\begin{equation}
m,n\rightarrow\infty,\quad m/n=\mathrm{const\quad}\text{(uniform limit)}
\end{equation}
for ratios $q_{n,m}^{\pm}$, $r_{n,m}$ is the simplest version of the
large-size limit for EST [see eqs. (\ref{q-uniform-limit}),
(\ref{r-uniform-limit})]. However, when it comes to lattice tests of EST in 3D
$Z_{2}$ gauge theory (considered below), another limit is easier for MC
simulations
\begin{equation}
n\rightarrow\infty,\quad m=1\mathrm{\quad}\text{(}m=1\text{ limit)}\,.
\end{equation}
This new $m=1$ limit can be also described by EST but one needs some care
about the order counting for the one-loop power correction. This power
correction is computed only for rectangles \cite{BCVG-2010},
\cite{Billo:2011fd} but the structure of this correction (order counting and
scaling properties) is under control for any polygon.

In case of hexagons $H_{n,m}$ the transition from the one-loop EST expression
for Wilson loops $W_{\text{EST}}^{\text{1-loop}}\left(  H_{n,m}\right)  $ to
the two-loop EST result $W_{\text{EST}}^{\text{2-loop}}\left(  H_{n,m}\right)
$ is given by
\begin{equation}
W\left(  H_{n,m}\right)  =W_{\text{EST}}^{\text{1-loop}}\left(  H_{n,m}
\right)  \left[  1+\frac{1}{n^{2}}f\left(  \frac{m}{n}\right)  +\ldots\right]
\,.
\end{equation}
Here $f\left(  x\right)  $ is an unknown function. Within this two-loop
accuracy we find from eq.~(\ref{r-n-m-def})
\begin{equation}
r_{n,1}\overset{n\rightarrow\infty}{=}r_{n,1}^{\mathrm{EST1}}\frac{\sqrt
{\left[  1+\frac{1}{n^{2}}f\left(  \frac{1}{n}\right)  \right]  \left[
1+\frac{1}{n^{2}}f\left(  -\frac{1}{n}\right)  \right]  }}{1+\frac{1}{n^{2}
}f\left(  0\right)  }
\end{equation}
where
\begin{equation}
r_{n,1}^{\mathrm{EST1}}=\left\{  \frac{\mathrm{Det}\left[  -\Delta_{\zeta
}\left(  H_{n,0}\right)  \right]  }{\sqrt{\mathrm{Det}\left[  -\Delta_{\zeta
}\left(  H_{n,1}\right)  \right]  \mathrm{Det}\left[  -\Delta_{\zeta}\left(
H_{n,-1}\right)  \right]  }}\right\}  ^{D-2}\,. \label{r-EST1-n-1}
\end{equation}
Here
\begin{equation}
\frac{\sqrt{\left[  1+\frac{1}{n^{2}}f\left(  \frac{1}{n}\right)  \right]
\left[  1+\frac{1}{n^{2}}f\left(  -\frac{1}{n}\right)  \right]  }}
{1+\frac{1}{n^{2}}f\left(  0\right)  }\overset{n\rightarrow\infty}
{=}1+\frac{1}{2n^{4}}f^{\prime\prime}\left(  0\right)  +O\left(
n^{-6}\right)  \,.
\end{equation}
Hence
\begin{equation}
r_{n,1}\overset{n\rightarrow\infty}{=}r_{n,1}^{\mathrm{EST1}}\left[
1+O\left(  n^{-4}\right)  \right]  \,.
\end{equation}
A similar argument shows that
\begin{equation}
\frac{\mathrm{Det}\left[  -\Delta_{\zeta}\left(  H_{n,0}\right)  \right]
}{\sqrt{\mathrm{Det}\left[  -\Delta_{\zeta}\left(  H_{n,1}\right)  \right]
\mathrm{Det}\left[  -\Delta_{\zeta}\left(  H_{n,-1}\right)  \right]  }
}=1+O\left(  n^{-2}\right)  \,.
\end{equation}
We see that in the $m=1$ case EST\ predicts that
\begin{equation}
r_{n,1}^{\mathrm{EST1}}\overset{n\rightarrow\infty}{=}1+O\left(
n^{-2}\right)  \,, \label{r-n-1-order-counting-1}
\end{equation}
\begin{equation}
r_{n,1}\overset{n\rightarrow\infty}{=}1+O\left(  n^{-2}\right)  \,,
\end{equation}
\begin{equation}
\frac{r_{n,1}}{r_{n,1}^{\mathrm{EST1}}}\overset{n\rightarrow\infty}
{=}1+O\left(  n^{-4}\right)  \,. \label{EST-2-loop-correction}
\end{equation}
As a consequence,
\begin{equation}
\lim_{n\rightarrow\infty}\frac{\left|  r_{n,1}-r_{n,1}^{\mathrm{EST1}}\right|
}{\left|  r_{n,1}^{\mathrm{EST1}}-1\right|  }=0\,.
\label{r-n-1-order-counting-4}
\end{equation}

Repeating the same arguments for ratios $q_{n,1}^{\pm}$ (\ref{q-n-m-def}) and
defining
\begin{equation}
\left(  q_{n,1}^{\pm}\right)  ^{\mathrm{EST1}}=e^{\sigma n}\left[
\frac{\mathrm{Det}\left[  -\Delta_{\zeta}\left(  H_{n,\pm1}\right)  \right]
}{\mathrm{Det}\left[  -\Delta_{\zeta}\left(  H_{n,0}\right)  \right]
}\right]  ^{\pm(D-2)/2}\,, \label{q-n-1-pm-def}
\end{equation}
we arrive at
\begin{equation}
\frac{q_{n,1}^{\pm}}{\left(  q_{n,1}^{\pm}\right)  ^{\mathrm{EST1}}}
\overset{n\rightarrow\infty}{=}1+O\left(  n^{-3}\right)  \,,
\label{q-ratio-asymp-1}
\end{equation}
\begin{equation}
e^{-\sigma n}\left(  q_{n,1}^{\pm}\right)  ^{\mathrm{EST1}}=1+O\left(
n^{-2}\right)  \,. \label{q-ratio-asymp-2}
\end{equation}

\section{Comparison of lattice MC results with EST}

\label{section-lattice-results}

\setcounter{equation}{0} 

\subsection{Lattice parameters}

Our lattice MC\ computation of Wilson loops $W\left(  H_{n,m}\right)  $ for
hexagons $H_{n,m}$ (\ref{H-n-m-geometry}) in gauge theory $\mathbb{Z}_{2}$ was
done in the dual Ising model at
\begin{equation}
\beta_{I}=0.23\,. \label{beta-Ising-0-23}
\end{equation}
According to (\ref{duality-beta-1}) this corresponds to $\mathbb{Z}_{2}$
coupling
\begin{equation}
\beta_{\mathbb{Z}_{2}}=0.743543\,. \label{beta-Z2-used}
\end{equation}
These values of $\beta_{\mathbb{Z}_{2}}$ and $\beta_{I}$ deviate from the
critical point (\ref{beta-Ising-critical}) and (\ref{beta-Z2-critical})
towards the confinement phase of the $\mathbb{Z}_{2}$ gauge theory (which
corresponds to the spontaneously broken phase of Ising model).

Our choice of values (\ref{beta-Ising-critical}) and (\ref{beta-Z2-critical})
for MC\ simulation is a compromise between two constraints:

-- continuous limit $\beta_{\mathbb{Z}_{2}}\rightarrow\beta_{\mathbb{Z}_{2}
}^{\mathrm{crit}}$,

-- large-size limit of Wilson contours $C$.

From the formal theoretical point of view, the continuous limit $\beta
_{\mathbb{Z}_{2}}\rightarrow\beta_{\mathbb{Z}_{2}}^{\mathrm{crit}}$ must be
taken first and only after that one can study the asymptotic large-size
expansion of Wilson loops. In practical MC\ simulations the choice of relevant
values of $\beta$ and loop-sizes for the comparison of EST is a subtle
problem. Our choice of $\beta_{I}$ value (\ref{beta-Ising-0-23}) is motivated
by several factors:

1) Agreement of other MC simulations using $\beta_{I}$ value
(\ref{beta-Ising-0-23}) and ``worse'' values of $\beta_{I}$ with EST, see e.g. 
ref.~\cite{BCGMP-12}.

2) In case of the large-size limit for polygonal Wilson loops the accuracy of
EST is controlled by the size of single sides of the polygon and not by the
overall size of the polygon. Hence one has to work with rather large polygons.
Therefore it makes sense to sacrifice the ``quality'' of $\beta_{I}$ and to
invest computer resources in the computation of larger Wilson loops.

3) String tension $\sigma$ of the $\mathbb{Z}_{2}$ model (coinciding with the
surface tension of Ising model) at point $\beta_{I}$ (\ref{beta-Ising-0-23}),
is known with a high precision \cite{BCGMP-12}:

\begin{equation}
\sigma=0.0228068(15)\,. \label{sigma-external}
\end{equation}
This high-precision value of $\sigma$ is used in our comparison of MC results
with EST\ in sec.~\ref{method-2-section}.

We compute Wilson loops for polygons $H_{n,m}$ (\ref{H-n-m-geometry}) with
\begin{equation}
m=0,\pm1
\end{equation}
in the large $n$ limit. The following values of $n$ were used in our MC computation:

\begin{equation}
n=8,16,24,32\,.
\end{equation}
Lattice sizes used in the MC computation of Wilson loops $W\left(
H_{n,m}\right)  $ are listed in Table~1.

\begin{table}[ptb]
\begin{tabular}
[c]{|l|l|l|l|l|}\hline
Size parameter $n$ of hexagon $H_{n,m}$ & $8$ & $16$ & $24$ & $32$\\\hline
Lattice size & $96^{3}$ & $96^{3}$ & $120^{3}$ & $180^{3}$\\\hline
\end{tabular}
\caption{3D lattice sizes used for the MC computation of ratios of
hexagon Wilson loops $W\left(  H_{n,m}\right)$.}
\label{table-1}
\end{table}

The technical implementation of MC simulations uses standard optimization
methods which are well described in literature \cite{Billo:2011fd}, \cite{Caselle:2002ah}.
The HMC algorithm (sec.~\ref{hierarchical-algorith-section}) is crucial for reaching a rather high
accuracy of MC\ results.

\subsection{Two methods for the analysis of lattice MC data}

One can compare lattice MC results with EST for large Wilson loops using two
methods. Method 1 deals with universal predictions of EST which are formulated
in terms of ratios of Wilson loops $r_{n,1}$ (\ref{r-n-m-def}) without using
string tension $\sigma$. Method 2 uses high-precision value of string tension
(\ref{sigma-external}) and deals with quantities $q_{n,1}^{\pm}$
(\ref{q-n-m-def}) involving both ratios of Wilson loops and string tension
$\sigma$.

An advantage of method 1 is its simplicity which allows for a fast but
incomplete comparison of lattice MC results with EST\ predictions. Ratios
$r_{n,1}$ used in method 1 have a universal large-size one-loop asymptotic
behavior independent of the gauge group. However,

1) $q_{n,1}^{\pm}$ are primary MC computed quantities, ratios $r_{n,1}$ are
computed via MC results for $q_{n,1}^{\pm}$ using eq.~(\ref{r-n-m-def}).

2) EST provides predictions not only for $r_{n,1}$ but also for $q_{n,1}^{\pm
}$ (if one knows string tension $\sigma$ from external MC\ results).

3) When one computes $r_{n,1}$ via $q_{n,1}^{\pm}$ (\ref{r-n-m-def}), one has
a certain information loss. Indeed, according to (\ref{q-n-1-pm-def}) --
(\ref{q-ratio-asymp-2}) at large $n$

\begin{equation}
\left|  q_{n,1}^{+}-q_{n,1}^{-}\right|  \ll\left|  1-q_{n,1}^{\pm}\right|
\end{equation}
so that in ratio $r_{n,1}=q_{n,1}^{-}/q_{n,1}^{+}$

-- significant information cancels,

-- statistical errors increase.

To summarize, method 1 is simple and transparent but incomplete in comparison
with more thorough method 2.

\subsection{Method 1: universal predictions of EST}

\label{method-1-section}

Method 1 is based on ratios of Wilson loops $r_{n,1}$. Theoretically ratios
$r_{n,1}$ are defined by eq.~(\ref{r-n-m-def})
\begin{equation}
r_{n,1}=\frac{W\left(  H_{n,-1}\right)  W\left(  H_{n,1}\right)  }{\left[
W\left(  H_{n,0}\right)  \right]  ^{2}}\,.
\end{equation}
Here $W\left(  H_{n,m}\right)  $ is Wilson loop for hexagon $H_{n,m}$ whose
geometry is given by eq.~(\ref{H-n-m-geometry}), see Fig. 1b. We use notation
$r_{n,1}^{\mathrm{MC}}$ for lattice MC data and $r_{n,1}^{\mathrm{EST1}}$ for
one-loop EST\ results. In lattice MC computation in Ising model based on
HMC algorithm (see sec. \ref{hierarchical-algorith-section}),
MC values $r_{n,1}^{\mathrm{MC}}$ are computed using decomposition
\begin{equation}
r_{n,1}^{\mathrm{MC}}=\frac{\left(  q_{n,1}^{-}\right)  ^{\mathrm{MC}}
}{\left(  q_{n,1}^{+}\right)  ^{\mathrm{MC}}}\,, \label{r-1-MC-via-q-1-MC}
\end{equation}

\begin{equation}
\left(  q_{n,1}^{-}\right)  ^{\mathrm{MC}}=\left[  \frac{W\left(
H_{n,-1}\right)  }{W\left(  H_{n,0}\right)  }\right]  ^{\mathrm{MC}}\,,
\end{equation}
\begin{equation}
\left(  q_{n,1}^{+}\right)  ^{\mathrm{MC}}=\left[  \frac{W\left(
H_{n,0}\right)  }{W\left(  H_{n,1}\right)  }\right]  ^{\mathrm{MC}}\,.
\end{equation}
In HMC method ratios $\left(  q_{n,1}^{\pm}\right)  ^{\mathrm{MC}}$
are computed directly without computing single Wilson loops $W\left(
H_{n,m}\right)  $. The resulting values $r_{n,1}^{\mathrm{MC}}$ are listed in
Table~2. The underlying values of $\left(  q_{n,1}^{\pm}\right)  ^{\mathrm{MC}}$
can be found in Table~3.

The one-loop EST prediction for $r_{n,1}$ is given by eq.~(\ref{r-EST1-n-1})
taken at $D=3$
\begin{equation}
r_{n,1}^{\mathrm{EST1}}=\frac{\mathrm{Det}\left[  -\Delta_{\zeta}\left(
H_{n,0}\right)  \right]  }{\sqrt{\mathrm{Det}\left[  -\Delta_{\zeta}\left(
H_{n,1}\right)  \right]  \mathrm{Det}\left[  -\Delta_{\zeta}\left(
H_{n,-1}\right)  \right]  }}\,. \label{r-1-EST}
\end{equation}
Table~2 lists numerical values of $r_{n,1}^{\mathrm{EST1}}$ computed using
numerical values Laplace determinants from Table~4.

In order to check the agreement of EST ratios $r_{n,1}^{\mathrm{EST1}}$ with
corresponding lattice MC\ ratios $r_{n,1}^{\mathrm{MC}}$, one should keep in
mind the accuracy of EST asymptotic formulas (\ref{r-n-1-order-counting-1}) --
(\ref{r-n-1-order-counting-4}). The data presented in Table~2 definitely agree
with the main prediction of EST (\ref{r-n-1-order-counting-4})
\begin{equation}
\lim_{n\rightarrow\infty}\frac{\left|  r_{n,1}-r_{n,1}^{\mathrm{EST1}}\right|
}{\left|  r_{n,1}^{\mathrm{EST1}}-1\right|  }=0\,.
\end{equation}
However, the confirmation of the more subtle limit testing the scaling of the
two-loop EST correction (\ref{EST-2-loop-correction})
\begin{equation}
d\left(  n\right)  =n^{4}\left(  1-\frac{r_{n,1}^{\mathrm{MC}}}{r_{n,1}
^{\mathrm{EST1}}}\right)  \,,
\end{equation}
\begin{equation}
\lim_{n\rightarrow\infty}d\left(  n\right)  =\text{finite}\neq0
\label{EST-2-loop-correction-again}
\end{equation}
is less reliable because

1) at $n=8$ asymptotic formulas of EST may have large contributions from
higher orders,

2) at $n=32$ the MC\ data have a too large statistical error.

Therefore one gets a confirmation for scaling
(\ref{EST-2-loop-correction-again}) of the two-loop EST correction only from
the agreement of $n=16$ and $n=24$ values for $n^{4}\left(  r_{n,1}
/r_{n,1}^{\mathrm{EST1}}-1\right)  $.

\noindent
\begin{table}[ptb]
\begin{tabular}
[c]{|l|l|l|l|l|l|}\hline
$n$ & $r_{n,1}^{\mathrm{EST1}}\vphantom{\Biggl|}  $ & $r_{n,1}^{\mathrm{MC}}$
& $\frac{r_{n,1}^{\mathrm{MC}}}{r_{n,1}^{\mathrm{EST1}}}-1$ & $\frac{r_{n,1}
^{\mathrm{MC}}-r_{n,1}^{\mathrm{EST1}}}{r_{n,1}^{\mathrm{EST1}}-1}$ &
\rule{0pt}{1.8em} $d\left(  n\right)  $\\\hline
8 & 1.00312545 & 1.001703(13) & $-0.001418(13)$ & $-0.455(4)$ & $5.808(53)$
\\\hline
16 & 1.00076881 & 1.0006323(55) & $-0.0001364(55)$ & $-0.1776(72)$ &
$8.94(36)$\\\hline
24 & 1.00034068 & 1.0003190(44) & $-0.0000216(44)$ & $-0.063(13)$ &
$7.2\pm1.5$\\\hline
32 & 1.00019143 & 1.000196(18) & $0.000004(18)$ & $0.023(92)$ & $-5\pm
19$\\\hline
\end{tabular}
\caption{Comparison of lattice MC\ results for triple hexagon Wilson
loop ratios $r_{n,1}^{\mathrm{MC}}$ with EST asymptotic expression
$r_{n,1}^{\mathrm{EST1}}$.}
\label{table-2}
\end{table}

\subsection{Method 2: analysis using the value of the string tension}

\label{method-2-section}

Method 2 is based on quantities $q_{n,1}^{\pm}$ which are defined by eq.
(\ref{q-n-m-def})
\begin{equation}
q_{n,1}^{\pm}=\left[  \frac{W\left(  H_{n,0}\right)  }{W\left(  H_{n,\pm
1}\right)  }\right]  ^{\pm1}\,.
\end{equation}
EST predictions for these ratios are described by eqs. (\ref{q-n-1-pm-def}) --
(\ref{q-ratio-asymp-2}). We introduce notation:
\begin{equation}
\left(  q_{n,1}^{+}\right)  ^{\mathrm{EST0}}=\left(  q_{n,1}^{-}\right)
^{\mathrm{EST0}}=e^{\sigma n}\,. \label{q-EST-0-def}
\end{equation}
We use the value of string tension $\sigma$ (\ref{sigma-external}).
Quantities $\left(  q_{n,1}^{\pm}\right)  ^{\mathrm{EST1}}$ are defined by eq.
(\ref{q-n-1-pm-def}). According to (\ref{q-ratio-asymp-1})
\begin{equation}
\frac{q_{n,1}^{\pm}}{\left(  q_{n,1}^{\pm}\right)  ^{\mathrm{EST1}}}
-1\overset{n\rightarrow\infty}{=}O\left(  n^{-3}\right)  \,. \label{q-q-EST1}
\end{equation}

In Table~3 we list

1) lattice MC\ results for $\left(  q_{n,1}^{\pm}\right)  ^{\mathrm{MC}}$,

2) values for quantity $\left(  q_{n,1}^{\pm}\right)  ^{\mathrm{EST0}}$
(\ref{q-EST-0-def}) using the value of string tension $\sigma$
(\ref{sigma-external}),

3) values for $\left(  q_{n,1}^{\pm}\right)  ^{\mathrm{EST1}}$
(\ref{q-n-1-pm-def}) computed using $\sigma$ (\ref{sigma-external}) and
Laplace determinants listed in Table~4.

The data presented in Table~3 show a fast decay of
\begin{equation}
\frac{\left(  q_{n,1}^{\alpha}\right)  ^{\mathrm{MC}}}{\left(  q_{n,1}
^{\alpha}\right)  ^{\mathrm{EST1}}}-1
\end{equation}
at large $n$. For all values $n=8,16,24,32$ we have
\[
\frac{\left|  \left(  q_{n}^{\alpha}\right)  ^{\mathrm{EST1}}-\left(
q_{n}^{\alpha}\right)  ^{\mathrm{MC}}\right|  }{\left|  \left(  q_{n}^{\alpha
}\right)  ^{\mathrm{EST0}}-\left(  q_{n}^{\alpha}\right)  ^{\mathrm{MC}
}\right|  }\ll1\,.
\]
However, the $O\left(  n^{-3}\right)  $ behavior of EST prediction
(\ref{q-q-EST1}) cannot be confirmed because of the insufficient accuracy of
lattice MC results for $n=24$ and $n=32$.

\begin{table}[ptb]
\begin{tabular}
[c]{|l|l|l|l|l|l|}\hline
$n$ & $\alpha$ & $\left(  q_{n,1}^{\alpha}\right)  ^{\mathrm{EST0}}$ &
$\left(  q_{n,1}^{\alpha}\right)  ^{\mathrm{EST1}}$ & $\left(  q_{n,1}
^{\alpha}\right)  ^{\mathrm{MC}}\vphantom{\Biggl|}  $ & $\frac{\left(
q_{n,1}^{\alpha}\right)  ^{\mathrm{MC}}}{\left(  q_{n,1}^{\alpha}\right)
^{\mathrm{EST1}}}-1$\\\hline
8 & $-$ & 1.200159(14) & 1.221022(15) & 1.221009(11) & $-0.000011(15)$\\\hline
8 & $+$ & 1.200159(14) & 1.217218(15) & 1.218933(11) & $0.001409(15)$\\\hline
16 & $-$ & 1.440383(35) & 1.452138(35) & 1.4521592(55) & $0.000014(24)$
\\\hline
16 & $+$ & 1.440383(35) & 1.451023(35) & 1.4512416(58) & $0.000151(24)$
\\\hline
24 & $-$ & 1.728689(62) & 1.737913(63) & 1.7376523(52) & $-0.000150(36)$
\\\hline
24 & $+$ & 1.728689(62) & 1.737321(63) & 1.7370981(56) & $-0.000128(36)$
\\\hline
32 & $-$ & 2.07470(10) & 2.08293(10) & 2.082510(27) & $-0.000200(50)$\\\hline
32 & $+$ & 2.07470(10) & 2.08253(10) & 2.082102(25) & $-0.000204(49)$\\\hline
\end{tabular}
\caption{Comparison of lattice MC\ results for double ratios $\left(
q_{n,1}^{\pm}\right)  ^{\mathrm{MC}}$ of hexagon Wilson loops with EST
asymptotic leading-order values $\left(  q_{n,1}^{\alpha}\right)
^{\mathrm{EST0}}$ and with one-loop results $\left(  q_{n,1}^{\alpha}\right)
^{\mathrm{EST1}}$.}
\label{table-3}
\end{table}

\section{Conclusions}

\setcounter{equation}{0} 

To summarize, our lattice MC results for large hexagon Wilson loops are in
a perfect agreement with EST. The achieved accuracy of MC\ simulations is
sufficient to confirm the one-loop contribution of the Laplace determinants.
The situation is less reliable for the two-loop correction. In the absence of
the full theoretical expression for this correction we could test only its
asymptotic scaling behavior which agrees with MC results but the accuracy of
this test is not sufficient for making final decisive conclusions.

\textbf{Acknowledgments}

My gratitude comes late to my senior friends who passed away: W.~Bathelt,
D.I.~Diakonov, K.~Goeke, M.~I.~Polykarpov and N.~G.~Uraltsev. I am grateful
for discussions and for a support to E.~T.~Akhmedov, A.~A.~Andrianov, 
E.~N.~Antonov, N.~V.~Antonov, G.~S.~Danilov, P.~Druck, M.~I.~Eides, E.~Epelbaum, 
N.~Gromov, H.~Dorn, V.~A.~Kudryavtsev, L.~N.~Lipatov, A.~S.~Losev, A.~D.~Mirlin,
M.~V.~Polyakov, N.~G.~Stefanis, T.~Takayanagi, S.~I.~Troyan, A.~V.~Yung and
many others. This work would not be possible without the help of 
V.~Yu.~Petrov. The MC simulations at 
Theoretical Physics Division  of
Petersburg Nuclear Physics Institute were
supported by Russian Science Foundation grant 14-22-00281. I also appreciate
the possibility to use computer facilities provided by Institute for
Theoretical Physics II of Ruhr University Bochum.

\renewcommand{\thesection}{\Alph{section}}  \setcounter{section}{1} 

\section*{Appendix. Computation of Laplace determinants}

\renewcommand{\theequation}{A.\arabic{equation}} 

\setcounter{equation}{0} 

\subsection{2D Laplace determinant for general polygons}

\subsubsection{Result of Aurell and Salomonson}

In this appendix we compute Laplace determinants for hexagons $H_{n,m}$
(\ref{H-n-m-geometry}) using the general results of ref.~\cite{AS-93-journal}.
In order to simplify the access to equations of ref.~\cite{AS-93-journal}, in
this appendix we try to be close to the original notation of ref.
\cite{AS-93-journal} (sometimes deviating from the notation used in the main
part of this paper).

In ref.~\cite{AS-93-journal} interior vertex angles $\theta_{\nu}$ of polygon
$P$ are parametrized by parameters $\beta_{\nu}$

\begin{align}
\theta_{\nu}  &  =\pi\left(  1-\beta_{\nu}\right)  \,,\label{theta-via-beta}\\
0  &  <\beta_{\nu}<1\,,\\
1  &  \leq\nu\leq M\,.
\end{align}
Sum rule (\ref{angle-sum-rule}) for the angles takes the form
\begin{equation}
\sum_{\nu=1}^{M}\beta_{\nu}=2\,\,.
\end{equation}

Eq.~(55) of ref.~\cite{AS-93-journal} defines Schwarz-Christoffel (SC) mapping
of the unit circle $\left|  u\right|  \leq1$ in complex $u$-plane to polygon
$P$ in the $z$-plane
\begin{equation}
\frac{dz}{du}=e^{\lambda_{0}}\prod_{\nu=1}^{M}\left(  u-e^{i\phi_{\nu}
}\right)  ^{-\beta_{\nu}}\,\,. \label{CS-circular}
\end{equation}
Here $e^{i\phi_{\nu}}$ are points on the boundary of the unit circle which are
mapped to vertices $z_{\nu}$ of polygon $P$. $\lambda_{0}$ is a real constant
controlling the size of the polygon.

The general result for the determinant of Laplace operator defined in polygon
$P$ with Dirichlet boundary conditions in $\zeta$-regularization
(\ref{Z-C-def}) -- (\ref{det-Z-prime}) is given by eq.~(62) of ref.
\cite{AS-93-journal}:
\begin{align}
-\ln\mathrm{Det}_{\zeta}\left[  -\Delta\left(  P\right)  \right]   &
=\sum\limits_{\mu=1}^{M}Z_{1-\beta_{\mu}}^{\prime}(0)+\frac{\lambda_{0}}
{12}\sum\limits_{\mu=1}^{M}\frac{\left(  2-\beta_{\mu}\right)  \beta_{\mu}
}{1-\beta_{\mu}}\nonumber\\
&  -\frac{1}{12}\sum\limits_{1\leq\mu\neq\nu\leq M}\frac{\beta_{\mu}\beta
_{\nu}}{1-\beta_{\mu}}\ln\left|  \left(  e^{i\phi_{\mu}}-e^{i\phi_{\nu}
}\right)  \right|  \label{AS-det-org}
\end{align}
where $Z_{\alpha}^{\prime}(0)$ is the original notation of ref.
\cite{AS-93-journal} for a certain function of parameter $\alpha$ (so that
$Z_{1-\beta_{\mu}}^{\prime}(0)$ should be understood as the value of this
function at $\alpha=1-\beta_{\mu}$). This function is defined by eqs. (51) --
(54) of ref.~\cite{AS-93-journal} and is discussed below in sec.
\ref{AS-2-section} of this paper.

The $\lambda_{0}$-dependent part on the RHS of eq.~(\ref{AS-det-org})
\begin{equation}
\frac{\lambda_{0}}{12}\sum\limits_{\mu=1}^{M}\frac{\left(  2-\beta_{\mu
}\right)  \beta_{\mu}}{1-\beta_{\mu}}=\frac{\lambda_{0}}{12}\sum
\limits_{\mu=1}^{M}\frac{\pi^{2}-\theta_{\mu}^{2}}{\pi\theta_{\mu}}
\end{equation}
controls the dependence of the Laplace determinant on the size of the polygon
and is in agreement with the scaling rule (\ref{Kac-dimension}),
(\ref{Gamma-cusp-string}), (\ref{log-Laplace-det-zeta-scaling}) for Laplace determinants.

\subsubsection{SC\ mapping in semiplane parametrization}

Paper \cite{AS-93-journal} uses SC mapping (\ref{CS-circular}) of the unit
circle to the polygon. One can map unit circle in the $u$-plane to the upper
semiplane of complex variable $\omega$ using linear-fractional transformation
\begin{equation}
\omega=i\frac{1-u}{1+u} \label{omega-via-u}
\end{equation}
with the inverse transformation
\begin{equation}
u=\frac{1+i\omega}{1-i\omega}\,\,. \label{u-via-omega}
\end{equation}

This maps points $e^{i\phi_{\nu}}$ on the boundary of the unit circle in the
$u$-plane to points $\omega_{\nu}$ on the real axis of the $\omega$-plane
\begin{equation}
\omega_{\nu}=\tan\left(  \frac{1}{2}\phi_{\nu}\right)  \,,
\label{omega-via-phi}
\end{equation}
\begin{equation}
e^{i\phi_{\nu}}=\frac{1+i\omega_{\nu}}{1-i\omega_{\nu}}\,.
\label{exp-i-phi-via-omega}
\end{equation}
Defining
\begin{equation}
\chi_{P}=\frac{1}{2}e^{\lambda_{0}}\left[  -i\exp\left(  \frac{i}{2}
\sum\limits_{\mu=1}^{M}\beta_{\mu}\phi_{\mu}\right)  \right]  \left[
\prod_{\mu=1}^{M}\left[  \cos\left(  \frac{1}{2}\phi_{\mu}\right)  \right]
^{-\beta_{\mu}}\right]\,,
\end{equation}
one finds that in terms of the $\omega$-plane parametrization, SC
transformation (\ref{CS-circular}) becomes
\begin{equation}
\frac{dz}{d\omega}=\chi_{P}\prod_{\nu=1}^{M}\left(  \omega-\omega_{\nu
}\right)  ^{-\beta_{\nu}}\,\,. \label{SC-semiplane}
\end{equation}
The upper semiplane
\begin{equation}
\mathrm{Im}\,\mathrm{\,}\omega\geq0
\end{equation}
is mapped to the polygon in the complex $z$-plane. Points $\omega_{\nu}$ on
the real axis of the $\omega$ plane are mapped to vertices $z_{\nu}$ of the
polygon in the $z$ plane. The side lengths $\left|  z_{\mu+1}-z_{\mu}\right|
$ of the polygon are given by integrals
\begin{equation}
\left|  z_{\mu+1}-z_{\mu}\right|  =\left|  \chi_{P}\right|  \int_{\omega_{\mu
}}^{\omega_{\mu+1}}d\omega\prod\limits_{\nu=1}^{M}\left|  \omega-\omega_{\nu
}\right|  ^{-\beta_{\nu}}\,. \label{CS-side-again}
\end{equation}
$\,$

In terms of this semiplane version of SC transformation, Laplace determinant
(\ref{AS-det-org}) becomes
\begin{equation}
\ln\mathrm{Det}_{\zeta}\left[  -\Delta\left(  P\right)  \right]  =\left\{
\ln\mathrm{Det}_{\zeta}\left[  -\Delta\left(  P\right)  \right]  \right\}
^{\left(  1\right)  }+\left\{  \ln\mathrm{Det}_{\zeta}\left[  -\Delta\left(
P\right)  \right]  \right\}  ^{\left(  2\right)  }
\label{Z-P-prime-semiplane-again}
\end{equation}
where
\begin{equation}
\left\{  \ln\mathrm{Det}_{\zeta}\left[  -\Delta\left(  P\right)  \right]
\right\}  ^{\left(  1\right)  }=\frac{1}{12}\sum\limits_{1\leq\mu\neq\nu\leq
M}\frac{\beta_{\mu}\beta_{\nu}}{1-\beta_{\mu}}\ln\left|  \frac{\omega_{\mu
}-\omega_{\nu}}{\chi_{P}}\right|  \,, \label{log-det-AS-part-1}
\end{equation}
\begin{equation}
\left\{  \ln\mathrm{Det}_{\zeta}\left[  -\Delta\left(  P\right)  \right]
\right\}  ^{\left(  2\right)  }=-\sum\limits_{\mu=1}^{M}Z_{1-\beta_{\mu}
}^{\prime}(0)\,. \label{log-det-AS-part-2}
\end{equation}

\subsubsection{Contribution $\left\{  \ln\mathrm{Det}_{\zeta}\left[
-\Delta\left(  P\right)  \right]  \right\}  ^{\left(  2\right)  }$}

\label{AS-2-section}

Contribution $\left\{  \ln\mathrm{Det}_{\zeta}\left[  -\Delta\left(  P\right)
\right]  \right\}  ^{\left(  2\right)  }$ (\ref{log-det-AS-part-2}) depends
only on the angles of the polygon and is additive. Therefore this contribution
cancels in any combinations of Wilson loops obeying vertex balance condition
(\ref{balance-angles-0}). From this point of view, term $\left\{
\ln\mathrm{Det}_{\zeta}\left[  -\Delta\left(  P\right)  \right]  \right\}
^{\left(  2\right)  }$ is irrelevant for our \emph{final} aims. On the other
hand, our \emph{intermediate} results deal with single Laplace determinants. In
particular, numerical values listed for Laplace determinants in Table~4
contain the contribution $\left\{  \ln\mathrm{Det}_{\zeta}\left[
-\Delta\left(  P\right)  \right]  \right\}  ^{\left(  2\right)  }$
(\ref{log-det-AS-part-2}) which depends on function $Z_{\alpha}^{\prime}(0)$.

The computation of function $Z_{\alpha}^{\prime}(0)$ is an essential part of
ref.~\cite{AS-93-journal}. However, in lattice applications one usually
deals with polygons containing only angles $\pi/2$ and $3\pi/2$ corresponding
to $\beta_{\mu}=\pm1/2$ so that one needs only the values for $Z_{1/2}
^{\prime}(0)$ and $Z_{3/2}^{\prime}(0)$. Function $Z_{\alpha}^{\prime}(0)$ is
a smooth function of parameter $\alpha$ in the range $-1<\alpha<1$ which
allows for an integral representation. The integral can be computed
analytically when $\alpha$ is rational, i.e. $\alpha=p/q$ where $p,q$ are
integer. According to eqs. (102), (103) of ref.~\cite{AS-93-journal}
\begin{align}
Z_{1/q}^{\prime}(0)  &  =\frac{q-1}{4q}\ln\pi-\left(  \frac{q}{12}-\frac{1}
{4}+\frac{1}{6q}\right)  \ln2\nonumber\\
&  +\left(  \frac{1}{4}-\frac{1}{12q}\right)  \ln q+\sum_{s=1}^{q-1}\left(
\frac{1}{2}-\frac{s}{q}\right)  \ln\Gamma\left(  \frac{s}{q}\right)  \,,
\label{AS-103}
\end{align}
\begin{align}
Z_{p/q}^{\prime}(0)  &  =\frac{q-p}{4q}\ln\left(  2\pi\right)  +\frac{p^{2}
-q^{2}}{12pq}\ln2-\frac{1}{q}\left(  p-\frac{1}{p}\right)  \zeta^{\prime
}\left(  -1\right) \nonumber\\
&  -\frac{1}{12pq}\ln q+\left[  \frac{1}{4}+S\left(  q,p\right)  \right]
\ln\frac{q}{p}\nonumber\\
&  +\sum_{r=1}^{p-1}\left(  \frac{1}{2}-\frac{r}{p}\right)  \ln\Gamma\left(
\frac{R\left(  rq,p\right)  }{p}\right) \nonumber\\
&  +\sum_{s=1}^{q-1}\left(  \frac{1}{2}-\frac{s}{q}\right)  \ln\Gamma\left(
\frac{R\left(  sp,q\right)  }{q}\right)  \,. \label{AS-102}
\end{align}
Here $\zeta^{\prime}$ is the derivative of Riemann $\zeta$ function:
\begin{equation}
\zeta\left(  z\right)  =\sum_{n=1}^{\infty}n^{-z}\,,
\end{equation}
\begin{equation}
\zeta^{\prime}\left(  z\right)  =\frac{d\zeta\left(  z\right)  }{dz}\,.
\end{equation}
$S\left(  q,p\right)  $ is Dedekind sum:
\begin{equation}
S\left(  q,p\right)  =\frac{1}{p}\sum_{r=0}^{p-1}r\left[  \frac{R\left(
rq,p\right)  }{p}-\frac{1}{2}\right]  \label{Dedekind-sum-def}
\end{equation}
where $R\left(  n,p\right)  $ is the non-negative remainder of the division of
$n$ by $p$:
\begin{equation}
n=pm+R\left(  n,p\right)  \,,\quad0\leq R\left(  n,p\right)  \leq p-1\,.
\end{equation}
Keep in mind that the definition of Dedekind sum given in eq.~(98) of article
\cite{AS-93-journal} has a typo: one should replace $S\left(  p,q\right)  $ by
$S\left(  q,p\right)  $ on the LHS of \emph{that} equation like it is done in
eq.~(\ref{Dedekind-sum-def}) of \emph{this} article. After this correction one
easily derives from eq.~(\ref{AS-103})
\begin{equation}
Z_{1/2}^{\prime}(0)=\frac{1}{8}\ln\left(  2\pi\right)  +\frac{1}{12}\ln2
\end{equation}
and from eq.~(\ref{AS-102})
\begin{equation}
Z_{3/2}^{\prime}(0)=-\frac{1}{8}\ln\left(  \frac{\pi}{2}\right)  -\frac{7}
{36}\ln3-\frac{4}{3}\zeta^{\prime}\left(  -1\right)  +\frac{1}{6}
\ln\frac{\Gamma\left(  2/3\right)  }{\Gamma\left(  1/3\right)  }
\end{equation}

A\ lattice polygon with $M$ vertices contains $\left(  M-4\right)  /2$
vertices with angle $3\pi/2$ and $\left(  M+4\right)  /2$ vertices with angle
$\pi/2$ so that for this class of polygons we find from eq.
(\ref{log-det-AS-part-2})
\begin{gather}
\left\{  \ln\mathrm{Det}_{\zeta}\left[  -\Delta\left(  P\right)  \right]
\right\}  ^{\left(  2\right)  }=-\left(  \frac{M+4}{2}\right)  Z_{1/2}
^{\prime}(0)-\left(  \frac{M-4}{2}\right)  Z_{3/2}^{\prime}(0)\nonumber\\
=-\left(  \frac{M+4}{2}\right)  \left[  \frac{1}{8}\ln\left(  2\pi\right)
+\frac{1}{12}\ln2\right] \nonumber\\
-\left(  \frac{M-4}{2}\right)  \left[  -\frac{1}{8}\ln\left(  \frac{\pi}
{2}\right)  -\frac{7}{36}\ln3-\frac{4}{3}\zeta^{\prime}\left(  -1\right)
+\frac{1}{6}\ln\frac{\Gamma\left(  2/3\right)  }{\Gamma\left(  1/3\right)
}\right]  \,.
\end{gather}

In case of a hexagon with $M=6$ this reduces to
\begin{equation}
\left\{  \ln\mathrm{Det}_{\zeta}\left[  -\Delta\left(  P\right)  \right]
\right\}  ^{\left(  2\right)  }=-\frac{1}{2}\ln\pi-\frac{7}{6}\ln
2+\frac{7}{36}\ln3+\frac{4}{3}\zeta^{\prime}\left(  -1\right)  -\frac{1}{6}
\ln\frac{\Gamma\left(  2/3\right)  }{\Gamma\left(  1/3\right)  }\,.
\label{log-det-part-2-hexagon}
\end{equation}

\subsubsection{Contribution $\left\{  \ln\mathrm{Det}_{\zeta}\left[
-\Delta\left(  P\right)  \right]  \right\}  ^{\left(  1\right)  }$}

The rest of the problem is the computation of $\left\{  \ln\mathrm{Det}
_{\zeta}\left[  -\Delta\left(  P\right)  \right]  \right\}  ^{\left(
1\right)  }$ (\ref{log-det-AS-part-1}). This computation consists of two parts:

1) For a given polygon with vertices given by complex plane coordinates
$z_{\mu}$, solve SC eq.~(\ref{CS-side-again}) and find real parameters
$\omega_{\mu}$ and (generally complex) parameter $\chi_{P}$. This inverse
problem has many solutions because of the freedom of real linear fractional
transformations mapping the upper semiplane of complex $\omega$ to itself.

2) Using solution $\left\{  \omega_{\mu}\right\}$, one can compute the RHS of
expression (\ref{log-det-AS-part-1}) for $\left\{  \ln\mathrm{Det}_{\zeta
}\left[  -\Delta\left(  P\right)  \right]  \right\}  ^{\left(  1\right)  }$,
the result is invariant with respect to real linear fractional transformations
of $\left\{  \omega_{\mu}\right\}$.

Generally, equation (\ref{CS-side-again}) for $\left\{  \omega_{\mu}\right\}
$ can be solved only numerically. But in the case of hexagon $H_{n,0}$
(\ref{H-n-m-geometry}) the integrals on the RHS of eq.~(\ref{CS-side-again})
can be expressed via the hypergeometric function.

\subsection{Case of symmetric hexagon}

\subsubsection{Hexagon $P_{ab}$}

Let us consider hexagon $P_{ab}$ with parametrization (\ref{P-general})
\begin{equation}
P_{ab}=P\left(  a+b,\frac{\pi}{2},a+b,\frac{\pi}{2},a,\frac{\pi}
{2},b,\frac{3\pi}{2},b\frac{\pi}{2},a,\frac{\pi}{2}\right)  \,.
\label{a-b-sides-hexagon}
\end{equation}
In case $a=b=n$ this hexagon coincides with hexagon $H_{n,0}$ defined by eq.
(\ref{H-n-m-geometry}) and used in our MC computation:
\begin{equation}
P_{nn}=H_{n,0}\,. \label{P-a-b-H-n-0}
\end{equation}
But in case $a\neq b$ hexagons $P_{ab}$ are different from hexagons $H_{n,0}$
(\ref{H-n-m-geometry}). Hexagons $P_{ab}$ are distinguished by the reflection
symmetry with respect to the diagonal axis passing through $3\pi/2$ vertex of
this polygon. It is well known that combining reflection symmetries of polygons
with Schwarz symmetry principle, one can simplify SC\ integrals. In particular,
the reflection symmetry of hexagons $P_{ab}$ allows to express SC\ integral
(\ref{CS-side-again}) via the hypergeometric function. Below we compute
Laplace determinant for the polygon $P_{ab}$ for the general case $a\neq b$
although for the comparison of our lattice MC results with EST

1) only the case $a=b$ is relevant,

2) we also need Laplace determinants for polygons $H_{n,\pm1}$ which cannot be
reduced to the hypergeometric function.

\subsubsection{Inverse SC problem for hexagon $P_{ab}$}

Using the diagonal reflection symmetry of hexagon $P_{ab}$, we can choose
parameters $\omega_{\mu}$ in the following way
\begin{equation}
\omega_{1}<\omega_{2}<\omega_{3}<\omega_{4}<\omega_{5}\,,
\label{omega-k-L-hexagon-ordering}
\end{equation}

\begin{align}
\omega_{3}  &  =0\,,\label{omega-3}\\
-\omega_{2}  &  =\omega_{4}=1\,,\\
-\omega_{1}  &  =\omega_{5}=w>1\,,\label{omega-1-5}\\
\omega_{6}  &  \rightarrow\infty\,. \label{omega-6}
\end{align}
We choose $\omega_{6}$ to be SC mapped to the $3\pi/2$ vertex of the polygon.
Then angular parameters $\beta_{\mu}$ (\ref{theta-via-beta}) of our hexagon
$P_{ab}$ are

\begin{equation}
\beta_{1}=\beta_{2}=\beta_{3}=\beta_{4}=\beta_{5}=-\beta_{6}=\frac{1}{2}\,.
\label{beta-1-6}
\end{equation}

In order to keep the size of the polygon fixed in the limit $\omega
_{6}\rightarrow\infty$, we must take the limit
\begin{equation}
\chi_{P}\rightarrow0
\end{equation}
for parameter $\chi_{P}$ in eq.~(\ref{SC-semiplane}) and keep the combination
\begin{equation}
\chi_{P}\left|  \omega_{6}\right|  ^{1/2}\equiv\psi_{P}=\mathrm{const}
\label{chi-P-psi-P}
\end{equation}
fixed in the limit $\omega_{6}\rightarrow\infty$.

Then SC equations (\ref{CS-side-again}) reduce to
\begin{align}
\left|  \psi_{P}\right|  \int_{0}^{1}d\omega f\left(  \omega\right)   &
=a+b\,,\label{int-L-hexagon-1}\\
\left|  \psi_{P}\right|  \int_{1}^{w}d\omega f\left(  \omega\right)   &
=a\,,\label{int-L-hexagon-2}\\
\left|  \psi_{P}\right|  \int_{w}^{\infty}d\omega f\left(  \omega\right)   &
=b\,, \label{int-L-hexagon-3}
\end{align}
where
\begin{equation}
f\left(  \omega\right)  =\left|  \omega\left(  \omega^{2}-w^{2}\right)
\left(  \omega^{2}-1\right)  \right|  ^{-1/2}\,.
\end{equation}
We find from eqs. (\ref{int-L-hexagon-1}), (\ref{int-L-hexagon-3})
\begin{equation}
\frac{a+b}{b}=\frac{\int_{0}^{1}d\omega f\left(  \omega\right)  }{\int
_{w}^{\infty}d\omega f\left(  \omega\right)  }\,. \label{ratio-a-plus-b-b}
\end{equation}
These integrals can be expressed via the hypergeometric function $F\left(
\alpha,\beta,\gamma,z\right)  $:
\begin{align}
F\left(  \alpha,\beta,\gamma,z\right)   &  =\sum_{n=0}^{\infty}\frac{\left(
\alpha\right)  _{n}\left(  \beta\right)  _{n}}{\left(  \gamma\right)  _{n}
}\frac{z^{n}}{n!},\\
\left(  \alpha\right)  _{n}  &  =\frac{\Gamma\left(  \alpha+n\right)  }
{\Gamma\left(  \alpha\right)  }\,,
\end{align}
\begin{equation}
I_{1}=\int_{0}^{1}d\omega f\left(  \omega\right)  =\frac{1}{2w\sqrt{2\pi}
}\left[  \Gamma\left(  \frac{1}{4}\right)  \right]  ^{2}F\left(  \frac{1}
{4},\frac{1}{2},\frac{3}{4},w^{-2}\right)  \,, \label{I1-res}
\end{equation}
\begin{equation}
I_{2}=\int_{w}^{\infty}d\omega f\left(  \omega\right)  =\left(  \frac{2\pi}
{w}\right)  ^{3/2}\left[  \Gamma\left(  \frac{1}{4}\right)  \right]
^{-2}F\left(  \frac{3}{4},\frac{1}{2},\frac{5}{4},w^{-2}\right)  \,.
\end{equation}
Now eq.~(\ref{ratio-a-plus-b-b}) gives
\begin{equation}
\frac{a+b}{b}=\frac{1}{8\pi^{2}}\left[  \Gamma\left(  \frac{1}{4}\right)
\right]  ^{4}\,w^{1/2}\frac{\,F\left(  \frac{1}{4},\frac{1}{2},\frac{3}
{4},w^{-2}\right)  }{\,F\left(  \frac{3}{4},\frac{1}{2},\frac{5}{4}
,w^{-2}\right)  }\,. \label{L-hexagon-aspect-ratio}
\end{equation}

For given geometrical parameters $a,b$ of hexagon $P_{ab}$, eq.
(\ref{L-hexagon-aspect-ratio}) determines parameter $w>1$ which in turn
controls SC parameters $-\omega_{1}=\omega_{5}=w$ (\ref{omega-1-5}). Thus we
have reduced the inverse SC\ problem (\ref{CS-side-again}) of expressing
SC\ parameters $\omega_{k}$ via geometrical parameters $z_{k}$ to eq.
(\ref{L-hexagon-aspect-ratio}).

\subsubsection{Laplace determinant for hexagon $P_{ab}$}

In case of

-- hexagons with angular parameters $\beta_{\mu}$ (\ref{beta-1-6}),

-- SC parameter $\omega_{6}$ taken to infinity

\noindent the sum on the RHS of eq.~(\ref{log-det-AS-part-1}) becomes
\begin{align}
&  2\sum\limits_{1\leq\mu\neq\nu\leq6}\frac{\beta_{\mu}\beta_{\nu}}
{1-\beta_{\mu}}\ln\left|  \frac{\omega_{\mu}-\omega_{\nu}}{\chi_{P}}\right|
\nonumber\\
&  =\sum\limits_{1\leq\mu\neq\nu\leq5}\ln\left|  \frac{\omega_{\mu}
-\omega_{\nu}}{\chi_{P}}\right|  -\sum\limits_{1\leq\mu\leq5}\ln\left|
\frac{\omega_{\mu}-\omega_{6}}{\chi_{P}}\right|  -\frac{1}{3}\sum
\limits_{1\leq\nu\leq5}\ln\left|  \frac{\omega_{\nu}-\omega_{6}}{\chi_{P}
}\right| \nonumber\\
&  \overset{\omega_{6}\rightarrow0}{\rightarrow}\sum\limits_{1\leq\mu\neq
\nu\leq5}\ln\left|  \omega_{\mu}-\omega_{\nu}\right|  -\frac{20}{3}\ln\left|
\omega_{6}\chi_{P}^{2}\right|  \,\,.
\end{align}
Combining this with (\ref{chi-P-psi-P}), we find
\begin{equation}
2\sum\limits_{1\leq\mu\neq\nu\leq6}\frac{\beta_{\mu}\beta_{\nu}}{1-\beta_{\mu
}}\ln\left|  \frac{\omega_{\mu}-\omega_{\nu}}{\chi_{P}}\right|  =\sum
\limits_{1\leq\mu\neq\nu\leq5}\ln\left|  \omega_{\mu}-\omega_{\nu}\right|
-\frac{40}{3}\ln\left|  \psi_{P}\right|  \,.
\end{equation}
In case of hexagon $P_{ab}$ with parameters $\omega_{\mu}$ (\ref{omega-3}) --
(\ref{omega-6}) we have
\[
\sum\limits_{1\leq\mu\neq\nu\leq5}\ln\left|  \omega_{\mu}-\omega_{\nu}\right|
=\ln\left|  4w^{3}\left(  w^{2}-1\right)  ^{2}\right|  \,.
\]
Now we insert these results into eq.~(\ref{log-det-AS-part-1})
\begin{equation}
\left[  \ln\mathrm{Det}_{\zeta}\left(  -\Delta\left(  P_{ab}\right)  \right)
\right]  ^{\left(  1\right)  }=\frac{1}{6}\ln\left|  2w^{3/2}\left(
w^{2}-1\right)  \right|  -\frac{5}{9}\ln\left(  \left|  \psi_{P}\right|
\right)  \,.
\end{equation}
According to eq.~(\ref{int-L-hexagon-1})
\begin{equation}
\psi_{P}=\frac{a+b}{I_{1}}
\end{equation}
so that
\begin{equation}
\left[  \ln\mathrm{Det}_{\zeta}\left(  -\Delta\left(  P_{ab}\right)  \right)
\right]  ^{\left(  1\right)  }=-\frac{5}{9}\ln\left(  a+b\right)  +\frac{1}
{6}\ln\left|  2w^{3/2}\left(  w^{2}-1\right)  \right|  +\frac{5}{9}\ln
I_{1}\,.
\end{equation}
Inserting this result with $I_{1}$ given by (\ref{I1-res}) into eq.
(\ref{Z-P-prime-semiplane-again}) and using expression
(\ref{log-det-part-2-hexagon}) for $\left\{  \ln\mathrm{Det}_{\zeta}\left[
-\Delta\left(  P\right)  \right]  \right\}  ^{\left(  2\right)  }$, we find
\begin{align}
&  \ln\mathrm{Det}_{\zeta}\left[  -\Delta\left(  P_{ab}\right)  \right]
=-\frac{5}{9}\ln\left(  a+b\right) \\
&  -\frac{1}{2}\ln\pi-\frac{7}{6}\ln2+\frac{7}{36}\ln3+\frac{4}{3}
\zeta^{\prime}\left(  -1\right)  -\frac{1}{6}\ln\frac{\Gamma\left(
2/3\right)  }{\Gamma\left(  1/3\right)  }\nonumber\\
&  +\frac{1}{6}\ln\left|  2w^{3/2}\left(  w^{2}-1\right)  \right|
+\frac{5}{9}\ln\left\{  \frac{1}{2\sqrt{2\pi}w}\left[  \Gamma\left(
\frac{1}{4}\right)  \right]  ^{2}\,F\left(  \frac{1}{4},\frac{1}{2}
,\frac{3}{4},w^{-2}\right)  \right\}  \,. \label{log-det-P-a-b}
\end{align}
This result, taken together with eq.~(\ref{L-hexagon-aspect-ratio}) defining
parameter $w$, solves the problem of the Laplace determinant for symmetric
hexagon $P_{ab}$ (\ref{a-b-sides-hexagon}).

\subsubsection{Laplace determinant for hexagon $H_{n,0}$}

Setting
\begin{equation}
a=b=n
\end{equation}
in eq.~(\ref{log-det-P-a-b}) and using eq.~(\ref{P-a-b-H-n-0}), we find
\begin{align}
&  \ln\mathrm{Det}_{\zeta}\left[  -\Delta\left(  H_{n,0}\right)  \right]
=-\frac{5}{9}\ln n\nonumber\\
&  +\frac{1}{18}\left(  -46\ln2+5\ln3-17\ln\pi\right)  +\frac{4}{3}
\zeta^{\prime}\left(  -1\right)  +\frac{1}{3}\ln\Gamma\left(  \frac{1}
{3}\right)  \nonumber\\
&  -\frac{11}{36}\ln w_{0}+\frac{1}{6}\ln\left(  w_{0}^{2}-1\right)
+\frac{5}{9}\ln\left\{  \left[  \Gamma\left(  \frac{1}{4}\right)  \right]
^{2}\,F\left(  \frac{1}{4},\frac{1}{2},\frac{3}{4},w_{0}^{-2}\right)
\right\}  \label{log-g-det-H-n0-via-w0}
\end{align}
where $w_{0}>1$ is the solution of equation (\ref{L-hexagon-aspect-ratio})
with $a=b$
\begin{equation}
\frac{1}{16\pi^{2}}\left[  \Gamma\left(  \frac{1}{4}\right)  \right]
^{4}\,w_{0}^{1/2}\frac{\,F\left(  \frac{1}{4},\frac{1}{2},\frac{3}{4}
,w_{0}^{-2}\right)  }{\,F\left(  \frac{3}{4},\frac{1}{2},\frac{5}{4}
,w_{0}^{-2}\right)  }=1\,. \label{log-det-via-u-L-hexagon-sym-a-eq-b}
\end{equation}
The numerical solution of eq.~(\ref{log-det-via-u-L-hexagon-sym-a-eq-b}) is

\begin{equation}
w_{0}=1.154700538379251529018297561\ldots
\end{equation}
Now eq.~(\ref{log-g-det-H-n0-via-w0}) gives
\begin{equation}
\ln\mathrm{Det}_{\zeta}\left[  -\Delta\left(  H_{1,0}\right)  \right]
=-1.11547761205543217467\ldots\label{log-det-H-1-1-res-numerical}
\end{equation}
Using eq.~(\ref{log-g-det-H-n0-via-w0}), one can rederive relation
(\ref{log-det-H-n-n-via-log-det-H-1-1}) expressing $\ln\mathrm{Det}_{\zeta
}\left[  -\Delta\left(  H_{n,0}\right)  \right]  $ via $\ln\mathrm{Det}
_{\zeta}\left[  -\Delta\left(  H_{1,0}\right)  \right]  $.

\subsection{Numerical results for Laplace determinants}

In case of polygons $H_{n,\pm1}$, SC integrals (\ref{CS-side-again}) can be
reduced (after taking the limit $\omega_{6}\rightarrow\infty$) to integrals of
type
\begin{equation}
\int_{\omega_{i}}^{\omega_{j}}d\omega\left[  \prod_{k=1}^{5}\left(
\omega-\omega_{k}\right)  \right]  ^{-1/2}\,.
\end{equation}
The computation of Laplace determinants for $H_{n,\pm1}$ repeats the same
steps as in the case of $H_{n,0}=P_{nn}$ considered above but now one has to combine

-- the computation of SC\ integrals (\ref{CS-side-again}),

-- the solution of equations (\ref{CS-side-again}) with respect to $\omega_{k}$

\noindent in one numerical bundle. One can recommend book \cite{DTbook-2002}
and software packages developed by its authors: SCPACK\ (Fortran, L.N.
Trefethen) \cite{Trefethen-site} and SC Toolbox\ (MATLAB, T.A. Driscoll)
\cite{Driscoll-site}. Although a straightforward numerical approach is not a
problem for simple polygons like our hexagons $H_{n,m}$, these packages (in
principle, designed for more complicated polygons) may be helpful.

\begin{table}[ptb]
\begin{tabular}
[c]{|l|l|l|l|}\hline
$n$ & $\ln\mathrm{Det}_{\zeta}\left[  -\Delta\left(  H_{n,-1}\right)  \right]
$ & $\ln\mathrm{Det}_{\zeta}\left[  -\Delta\left(  H_{n,0}\right)  \right]  $
& $\ln\mathrm{Det}_{\zeta}\left[  -\Delta\left(  H_{n,1}\right)  \right]
$\\\hline
8 & $-2.30519045940400$ & $-2.27072291298867$ & $-2.24249651205329$\\\hline
16 & $-2.67206125015434$ & $-2.65580467996642$ & $-2.64108514509615$\\\hline
24 & $-2.89170650424248$ & $-2.88106307335984$ & $-2.87110088157089$\\\hline
32 & $-3.04879906577170$ & $-3.04088644694416$ & $-3.03335665452480$\\\hline
\end{tabular}
\caption{Numerical results for logarithms of Laplace determinants
$\mathrm{Det}_{\zeta}\left[  -\Delta\left(  H_{n,m}\right)  \right]  $ in
$\zeta$-regularization for hexagon regions $H_{n,m}$ with Dirichlet boundary condition.}
\label{table-4}
\end{table}

Our final numerical results for all Laplace determinants needed for the
comparison of our lattice MC\ results with EST, i.e. for $\mathrm{Det}_{\zeta
}\left[  -\Delta\left(  H_{n,m}\right)  \right]  $ with
\begin{align}
n  &  =8,16,24,32\,,\\
m  &  =0,\pm1
\end{align}
are listed in Table~4.

\end{document}